\documentclass[twocolumn]{aa}  
\usepackage{graphicx}
\usepackage{txfonts}

\usepackage{bm}
\renewcommand{\vec}[1]{\bm{#1}}

\usepackage[]{hyperref}
\begin{document}

   \title{The GRAVITY fringe tracker}

   \author{
   S. Lacour
             \inst{1,2,3}         \and 
   R. Dembet
             \inst{1,4}         \and 
R. Abuter
             \inst{4}         \and 
P. Fédou
             \inst{1}         \and 
G.~Perrin
             \inst{1}       \and 
\'E.~Choquet
             \inst{5}         \and 
O.~Pfuhl
             \inst{2}       \and 
 F.~Eisenhauer
             \inst{2}         \and  
 J.~Woillez
             \inst{4}         \and  
 F.~Cassaing
             \inst{6}         \and  
E.~Wieprecht
             \inst{2}         \and  
T.~Ott
             \inst{2}         \and  
E.~Wiezorrek
             \inst{2}         \and  
K.R.W.~Tristram 
             \inst{4}         \and 
B.~Wolff
             \inst{4}         \and 
A.~Ram\'irez
             \inst{4}         \and 
X.~Haubois
             \inst{4}         \and 
   K.~Perraut
             \inst{7}       \and 
  C.~Straubmeier
             \inst{8}       \and 
W.~Brandner
             \inst{9}       \and 
A.~Amorim
             \inst{10}      
          }

   \institute{
   LESIA, Observatoire de Paris, Universit\'e PSL, 
CNRS, Sorbonne Universit\'e, Univ. Paris Diderot, 
Sorbonne Paris Cit\'e, 5 place Jules Janssen, 92195 Meudon, France
 \and 
Max Planck Institute for extraterrestrial Physics,
Giessenbachstraße~1, 85748 Garching, Germany
 \and 
 Department of Physics, University of Cambridge, CB3 0HE Cambridge, United Kingdom
\and 
  European Southern Observatory, Karl-Schwarzschild-Straße 2, 85748
Garching, Germany
\and
Aix Marseille Univ, CNRS, CNES, LAM, Marseille, France
\and
ONERA/DOTA, Université Paris Saclay, F-92322 Châtillon, France
\and
Univ. Grenoble Alpes, CNRS, IPAG, 38000 Grenoble, France
\and
$1^{\rm st}$ Institute of Physics, University of Cologne,
Z\"ulpicher Straße 77, 50937 Cologne, Germany
\and
Max Planck Institute for Astronomy, K\"onigstuhl 17, 
69117 Heidelberg, Germany
\and
CENTRA - Centro de Astrofísica e Gravitação, IST, Universidade de Lisboa, Campo Grande,
1749-016 Lisboa, Portugal 
             }

   \date{Received September 15, 1996; accepted March 16, 1997}

 
  \abstract
   {The GRAVITY instrument was commissioned on the VLTI in  2016 and is now available to the astronomical community. It is the first optical interferometer capable of observing sources as faint as magnitude 19 in K band. This is possible through the fringe tracker, which compensates the differential piston based on measurements of a brighter off-axis astronomical reference source.}
   { The goal of this paper is to describe the main developments made in the context of the GRAVITY fringe tracker. This could serve as basis for future fringe-tracking systems.}
   {The paper therefore covers all aspects of the fringe tracker, from hardware to control software and on-sky observations. Special emphasis is placed on the interaction between the group-delay controller and the phase-delay controller. The group-delay control loop is a simple but robust integrator. The phase-delay controller is a state-space control loop based on an auto-regressive representation of the atmospheric and vibrational perturbations. A Kalman filter provides the best possible determination of the state of the system. }
   {The fringe tracker shows good tracking performance on sources with coherent K magnitudes of 11 on the Unit Telescopes (UTs) and 9.5 on the Auxiliary Telescopes (ATs). It can track fringes with a signal-to-noise ratio of 1.5 per detector integration time, limited by photon and background noises. During good seeing conditions, the optical path delay residuals on the ATs can be  as low as 75\,nm root mean square. The performance is limited to around 250\,nm on the UTs because of structural vibrations. }
   {}

   \keywords{Instrumentation: interferometers
   -- Techniques: high angular resolution               }

   \maketitle
%

\section{Introduction}

GRAVITY \citep{2017A&A...602A..94G} is an instrument used on the Very Large Telescope Interferometer (VLTI) situated at the Cerro Paranal Observatory. It can combine the light from four telescopes. These telescopes can either be the four Auxiliary Telescopes (ATs, with a primary mirror diameter of 1.8 meter) or the four Unit Telescopes (UTs, with a diameter of 8 meters). 
The specifications of the instrument were derived from the most demanding  science case, which was
 to observe microarcsecond displacements of the light source causing the flares of the supermassive black hole Sgr A* \citep[][and references herein]{2010RvMP...82.3121G}. Such astrometric measurements are possible with 100-meter baselines \citep{1992A&A...262..353S,2014A&A...567A..75L} and were recently demonstrated on-sky by  \citet{2018A&A...615L..15G, 2018A&A...618L..10G, 2018arXiv181111195G}.
 
  In its quiescent state, Sgr A* can become fainter than K=18 mag. Therefore, measuring its position reliably requires an integration time on the order of minutes. To enable such long integration times, it is important to correct the atmosphere  effects in real time. The higher-order atmospheric wavefront distortions are compensated for by an adaptive optics (AO) system. However,
   the AO systems do not sense, and therefore cannot correct, the differential phase between the telescopes. 
This is the role of the fringe tracker: a phase-referencing target (IRS\ 16C in the case of Sgr A*) is used as a guide star. In real time, the optical path differences (OPD) between each pair of telescopes are computed, and are used to control the displacement of mirrors on piezoelectric systems.
This is the counterpart of the AO system, but at interferometric scale.

To push the comparison a little farther: without fringe tracking, interferometry requires short integration times and deconvolution techniques.
This was the time of speckle imaging  \citep{1970A&A.....6...85L,Weigelt1977}, when using bispectrum and closure phases was a good but not really sensitive technique.
 With fringe tracking, optical interferometry enters a new age: 
long detector integration times (DIT up to 300\,s) give access to faint sources (Kmag of 19) and to the possibility of combining spectral resolution (up to 4000) with milliarcsecond spatial resolution. 
This is the historical equivalent to the emergence of adaptive optics: it enables a new level of science.

Fringe tracking is not new, however. Small observatories demonstrated the concept, with PTI and CHARA \citep{2006SPIE.6268E..3KB}. On the Keck Interferometer \citep{2013PASP..125.1226C}, comparable astrophysical objectives \citep{2012PASP..124...51W} pushed a similar development for phase referencing  \citep{2010PASP..122..795C}. Previous projects also existed at the VLTI: at first, the FINITO fringe tracker \citep{2008A&A...481..553L} was used in combination with the AMBER instrument  \citep{2007A&A...464....1P}. More recently, ESO developed the PRIMA fringe tracker \citep{2006SPIE.6268E..0UD}. 

However, unlike AO, fringe tracking was not yet mature, and many complex problems had to be investigated for GRAVITY. A first problem was how to deal with limited degrees
of freedom (the piston actuators) while many more optical path differences are measured \citep{2012A&A...541A..81M}. A second problem, which does not exist in AO, is that the phase signals are known only modulo $2\pi$. A third problem, partially addressed by the AO community \citep{2008OExpr..16...87P,2010JOSAA..27A.223P}, is how to set a correct state space control system that optimally uses a 
 Kalman filter to cancel the vibrations \citep{2014A&A...569A...2C}. 
A fourth difficulty is that both group-delay (GD) and phase-delay (PD) tracking need to be used in a control system to keep the best of both. 
The last hurdle of the project consisted of dealing with multiple closing baselines, some of them resolved. The GRAVITY fringe tracker is now the best and most modern instrument in the field of fringe tracking for optical interferometry. Below we describe the algorithms and mechanisms.

This paper builds upon earlier works from  \citet{Cassaing-p-08a}, \citet{Houairi-p-08a}, \citet{Lozi-p-11}, \citet{2012A&A...541A..81M}, and \citet{2014A&A...569A...2C}\footnote{presented during the Final Design Review as document {\em VLT-TRE-GRA-15882-6701}.}. The \citet{2012A&A...541A..81M}  paper theoretically describes modal control of the phase delay. The \citet{2014A&A...569A...2C} paper  simulates the expected performance of the Kalman controller. The present paper wraps up the series by presenting the final implementation on the VLTI. Section~\ref{sec:tech} is an overview of the technical implementation of the fringe tracker and is followed by Section~\ref{sec:estim}, where the basis of the fringe sensing is defined by the observables. The control algorithm is presented in three sections: Section~\ref{sec:control1} defines the operational modes, Section~\ref{sec:control2} presents the group-delay
controller, and Section~\ref{sec:control3} presents the phase-delay controller. Section~\ref{sec:results} gives examples and statistics of on-sky observations. Last, Section~\ref{sec:conclusion} concludes the paper by a discussion of possible improvements to increase the sensitivity and accuracy of the fringe tracker.

\section{Overview of the fringe-tracking system}
\label{sec:tech}

\subsection{Hardware}

   \begin{figure}
   \centering
   \includegraphics[width=0.45\textwidth]{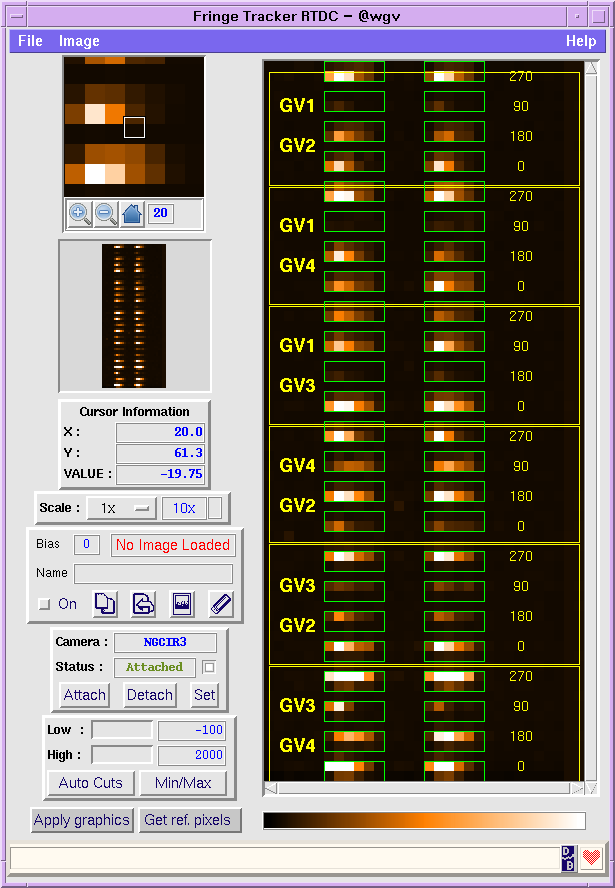}
   \caption{RTD of the SAPHIRA detector. The pixels in the green boxes are read by the fringe tracker and are used for tracking. They correspond to the six baselines, four ABCD outputs, two polarizations, and six wavelength channels. The names GV1 to GV4 correspond to the input beams. The values in yellow to the right correspond to the phase shift in degrees between the different ABCD outputs. }
              \label{Fig:RTD}%
    \end{figure}

GRAVITY is equipped with two beam combiners \citep{Karine08} that perform fringe tracking and scientific observations in the K band. GRAVITY has two main operational modes. In the on-axis mode, the light of one star is split 50:50, that is, equal portions of the flux go to the fringe tracking and to the scientific channel. In off-axis mode, the field is split into two with the help of a roof mirror: one of the two objects serves as fringe-tracking reference, while the scientific channel carries out longer integrations on the typically fainter science target \citep{2010SPIE.7734E..2AP,2012SPIE.8445E..1UP,2014SPIE.9146E..23P}.

The two beam combiners are based on silica-on-silicium integrated optics \citep{1999A&AS..138..135M}, optimized for $2\,\mu$m observations \citep{2010SPIE.7734E..30J}.
The beam combination scheme is pair-wise. Each telescope pair is combined using a static ABCD phase modulation. This means that for each of the six baselines, there are four outputs corresponding to a phase shift between the two beams of 0, $\pi/2$, $\pi,$ and $3\pi/2$ radians.  The total 24 outputs of the beam combiner can be seen on the real time display (RTD) of the instrument. The relevant pixels are delimited by the green rectangles in Fig.~\ref{Fig:RTD}. Within each rectangle, the flux is dispersed in six spectral channels. The 24 lines of rectangles correspond to the 24 outputs, while the two columns of rectangles correspond
to the two linear polarizations. 

The detector is a HgCdTe avalanche photodiode array
called SAPHIRA \citep{2016SPIE.9909E..12F}. It is running at 909Hz, 303Hz, or 97Hz. It can also run at either low or high gain. The high gain corresponds to a gain of $\gamma=7$\,ADU per photodetection with a typical readout noise below $\sigma_{\rm RON}=5$\,ADU ($\approx 0.7\,$e$^-$) per pixel. The low gain does not amplify the photodetections ($\gamma=0.5$\,ADU/e$^{-}$) and is only used for very bright targets (K magnitudes below 5 on the UTs).

The flux is processed by a first local control unit (LCU) that yields values of the observables. The LCU is an Artesyn MVME6100 using an MPC7457 PowerPC\textregistered processor \citep{2010SPIE.7740E..0TK}. The data are then transmitted to a second LCU by means of a distributed memory system called reflective memory network (RMN). 
This second LCU processes the observables 
to control four tip-tilt piston mirrors on piezoelectric actuators from Physik Instrument. Each actuator has its own position sensor, driven in closed loop. The cutoff frequency of the piezoelectric delay lines is then higher than $300$\,Hz, with a maximum optical path delay of $60\,\mu$m  \citep{2014SPIE.9146E..23P}. 

Real-time monitoring of the fringe tracking, including live display, is done on a separate Linux workstation connected to the LCUs through the RMN. This workstation processes the data, computes the best-fit control parameters (including the Kalman parameters), and updates the control parameters of the second LCU \citep{2016SPIE.9907E..21A}.

\subsection{Software}
\label{sec:software}

The two LCUs use the VxWorks operating systems. The computation is done using the so-called tools for advanced control (TAC) framework, which uses the standard C language environment. The TAC processing is triggered synchronously with the SAPHIRA, following the predefined frequency of the detector. 

The first LCU computes the necessary estimators for the controller: 
\begin{itemize}
\item Four flux values ($F_i$); see Sect.~\ref{sec:flux}.
\item Six phase delays ($\Phi_{i,j}$); see Sect.~\ref{sec:pd}. 
\item Six phase-delay variance ($Var(\Phi_{i,j})$); see Sect.~\ref{sec:snr}.
\item Six group delays ($\Psi_{i,j}$); see Sect.~\ref{sec:gd}.
\item Four closure phases ($\Theta_{i,j,k}$); see Sect.~\ref{sec:cp}.
\end{itemize}
Inside this LCU, only a few parameters can be changed: the number of DITs over which each of these quantities can be averaged. The default values are presented in section~\ref{sec:estim}.

\begin{figure*}
   \centering
   \includegraphics[width=\textwidth]{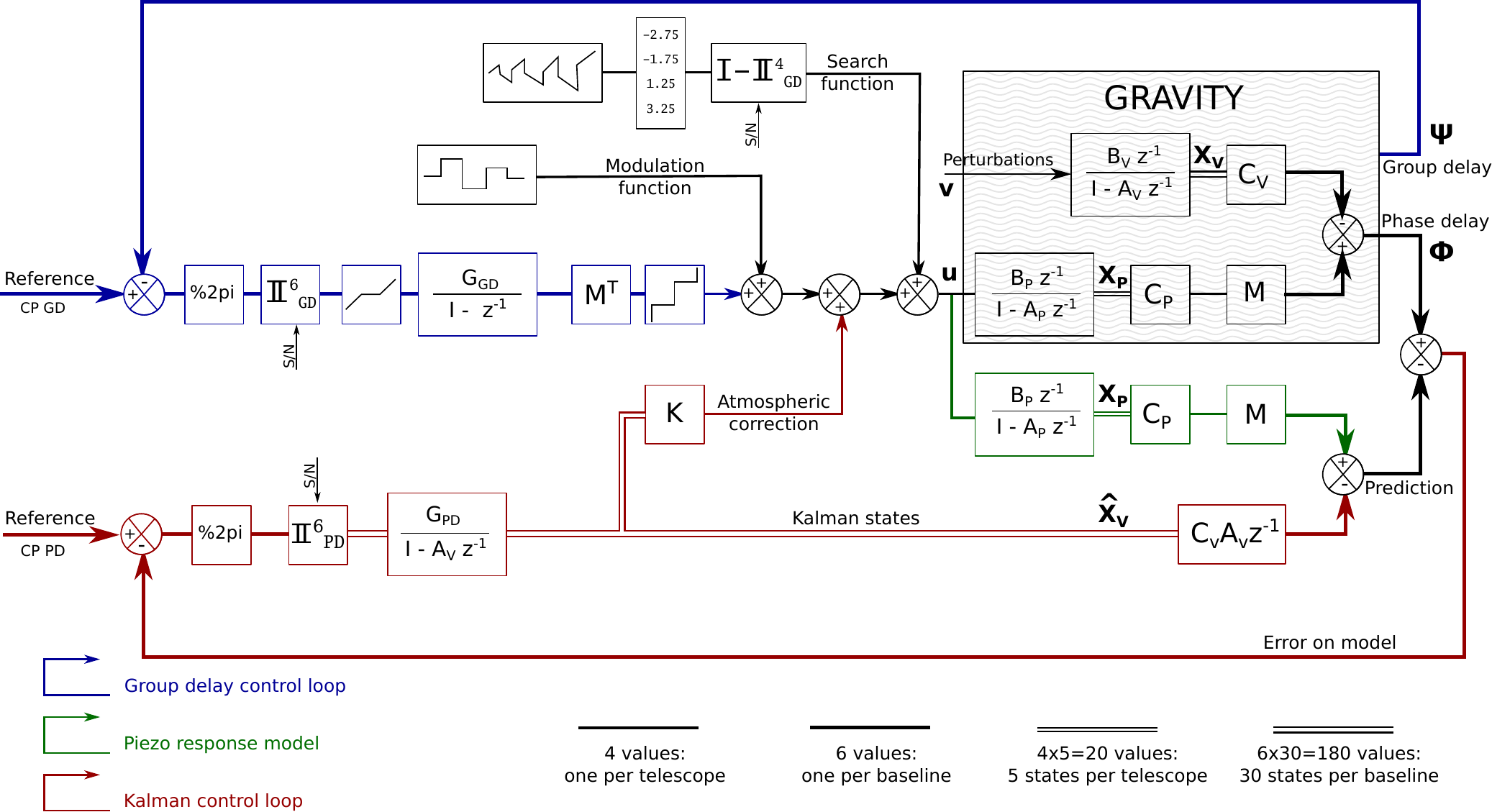}
   \caption{Block diagram of the GRAVITY fringe-tracking controller. The gray area corresponds to the GRAVITY open-loop transfer function, and the rest corresponds to the two controllers. The group-delay integrator controller is shown in blue, the phase-delay state controller in red, and the actuators predictive model is plotted in green. The group-delay controller is the main controller: it continues to track the fringes even if the instantaneous S/N is too low for phase-delay tracking. The phase-delay state controller is a closed-loop system that determines  the atmospheric perturbations $\hat { \vec X}_V$. A proportional controller ($\mathsf K$) corrects for the effect of the atmosphere. The last block of the group-delay controller, the quantization step, ensures that the group-delay control signal is always a multiple of $2\pi$: the change in OPD caused by the group-delay controller is not seen by the phase-delay controller.}
              \label{fig:control}%
    \end{figure*}

The second LCU is in charge of controlling the piezo-mirrors for adequate fringe tracking. Figure~\ref{fig:control} is a block diagram of the controller algorithm:
\begin{itemize}
\item The group-delay control loop, based on an integrator controller,  with a direct command to the piezo-actuator (in blue).
\item A feed-forward predictor, based on the action to the actuators, to increase the gain of the closed-loop system (in green).
\item The phase-delay Kalman controller (in red).
\item Two peripheral blocks for searching the fringes and adding a $\pi$ modulation to the phase delay.
\end{itemize}
The dual architecture of the controllers is made to obtain both sensitivity and accuracy. In case of high signal-to-noise ratio (S/N), the Kalman filter can determine and predict the state of both atmospheric and vibrational perturbation for the best possible correction.
In case of low S/N, the Kalman filter relies on its predictive model, which in the worst case, can be as simple as a constant value. In this case, the group-delay controller is still working efficiently and provides coherence instead of fringe tracking.
  
  The third and last software element is on the Linux workstation. It is a python script that runs every 5 seconds on the last 40 seconds of data calculated on the first LCU. It computes the parameters that are then used by the second LCU, which does the real time control. It includes the parameters for the predictive control and the best gain for the Kalman filter.
  
       
\section{Estimators of observables}
\label{sec:estim} 

\subsection{Visibility extraction}

The real-time processing load consists mostly of matrix multiplications. The pixel-to-visibility ($\mathsf{P2VM}$)  principle was first used for data reduction \citep{2007A&A...464...29T} on AMBER. AMBER used spatial modulation for fringe coding, but the formalism was subsequently adapted to work with ABCD beam combiners \citep{2008SPIE.7013E..16L}. Using a similar notation, the relation between the incoming electric field $E_i$ and the outgoing electric field $S_k$ can be expressed as
\begin{equation}
S_k=\sum_i T_{k,i} E_i
,\end{equation}
where $T_{k,i}$ is the complex transfer function from the input $i$ of the beam combiner to its output $k$. In the case of GRAVITY, $i_{\rm max}=4$ and $k_{\rm max}=24$. Averaged over the DIT, the flux is then equal to the average of its instantaneous intensity:
\begin{equation}
\langle |S_k|^2 \rangle_{\rm DIT}= \langle \left|\sum_n T_{k,i} E_i\right|^2  \rangle_{\rm DIT}
,\end{equation}
or alternatively, after decomposition:
\begin{equation}
\langle |S_k|^2 \rangle=\Re\left[ \sum_n |T_{k,i}|^2   \langle |E_i|^2 \rangle + 2 \sum_i\sum_{j>i}  \langle T_{k,i} T_{k,j}^*E_i E_j^* \rangle \right].
\label{eq:Sk}
\end{equation}
The temporal average of the electric field leads to two types of coherence losses. One depends on the optical path inside the beam combiner, the other on the spatial brightness distribution of the astrophysical object. The first, $C_{k,i,j}$, is intrinsic to the device and has to be calibrated. The second, $V_{i,j}$, is the reason why we built the interferometer. The relation with the mean electric field intensity is then approximated by the following equation:
\begin{equation}
\langle T_{k,i} T_{k,j}^*E_i E_j^* \rangle  =  |T_{k,i} T_{k,j}^*|  \  C_{k,i,j} \ \sqrt{\langle |E_i|^2 \rangle \langle |E_j|^2 \rangle}  \, V_{i,j}
.\end{equation}
Hence, Eq.~(\ref{eq:Sk}) can be written as a matrix product:
\begin{equation}
\left(
\begin {array}{c}
\langle |S_1|^2 \rangle \\
\vdots \\
\langle |S_{24}|^2 \rangle 
\end {array}
\right)
=  
\Re \left[ 
\mathsf{V2PM}
\cdot
\left(
\begin{array}{c}
\langle |E_1|^2\rangle \\
\langle |E_2|^2\rangle \\
\langle |E_3|^2\rangle \\
\langle |E_4|^2\rangle \\
\sqrt{\langle |E_1|^2 \rangle \langle |E_2|^2 \rangle}    V_{1,2}\\
\sqrt{\langle |E_1|^2 \rangle \langle |E_3|^2 \rangle}    V_{1,3}\\
\sqrt{\langle |E_1|^2 \rangle \langle |E_4|^2 \rangle}    V_{1,4}\\
\sqrt{\langle |E_2|^2 \rangle \langle |E_3|^2 \rangle}    V_{2,3}\\
\sqrt{\langle |E_2|^2 \rangle \langle |E_4|^2 \rangle}    V_{2,4}\\
\sqrt{\langle |E_3|^2 \rangle \langle |E_4|^2 \rangle}    V_{3,4}
\end{array}
\right)
\right]
\label{V2PMEq} 
,\end{equation}
where, using $^\top$ as the transpose operator:
\begin{equation}
\mathsf{V2PM}^\top
=  
\left(
\begin{array}{ccc}
 |T_{1,1}|^2 & \cdots &  |T_{24,1}|^2 \\
 \vdots & \ddots & \vdots \\
  |T_{1,4}|^2 & \cdots & |T_{24,4}|^2 \\
    |T_{1,1}T_{1,2}^*| C_{1,1,2} & \cdots &|T_{24,1}T_{24,2}^*| C_{24,1,2}\\
 \vdots & \ddots & \vdots \\
    |T_{1,3}T_{1,4}^*| C_{1,3,4}& \cdots &|T_{24,3}T_{24,4}^*| C_{24,3,4} \\
\end{array}
\right)\,.
\end{equation}
Everything related to the transfer function of the instrument is inside the 10 columns and 24 rows of the $\mathsf{V2PM}$ matrix. The $\mathsf{V2PM}$ is calibrated during daytime on the internal source of the instrument. It is regularly computed for verification, but has been proven to be very stable over several months. It is part of the calibration files that are needed by the first LCU (section~\ref{sec:software}).

The $\mathsf{P2VM}$ is the pseudo-inverse matrix of $\mathsf{V2PM}$. It can be obtained by splitting the $\mathsf{V2PM}$ into real and imaginary parts, and taking the inverse using a  
singular-value decomposition (SVD).
The $\mathsf{P2VM}$ is used to retrieve the astrophysical information from the flux observed on the pixels. This information consists of the flux $F$ and the coherent flux $\Gamma$. Both are computed from the pixel flux $q_{k,\lambda}$ and $\mathsf{P2VM}$ according to Eq~(\ref{eq:P2VM}):
\begin{equation}
\left(
\begin{array}{c}
F_{1,\lambda}\\
\vdots\\
F_{4,\lambda}\\
\Re [ \Gamma_{1,2,\lambda} ]\\
\vdots\\
 \Re [ \Gamma_{3,4,\lambda} ]\\
 \Im [ \Gamma_{1,2,\lambda} ]\\
\vdots\\
 \Im [ \Gamma_{3,4,\lambda} ]
\end{array}
\right) 
=
 \mathsf{P2VM}_\lambda \cdot \left(
\begin {array}{c}
q_{1,\lambda}\\
\vdots \\
q_{24,\lambda}
\end {array}
\right)
\label{eq:P2VM}
.\end{equation}
 In the above equation, we used a slightly different nomenclature, which is used hereafter. $q_{k,\lambda}=<|S_{k,\lambda}|^2>$ is the number of photons observed over one DIT of the fringe tracker on a given pixel.  $F_{i,\lambda}=<|E_{i,\lambda}|^2>$ is  the energy per DIT of the incoming beam $i$ at wavelength $\lambda$. Last, 
the complex coherent flux $\Gamma_{i,j,\lambda}$ corresponds to the visibility before normalization by the flux. It is obtained from the flux $F_{i,\lambda}$ and the visibilities $V_{i,j,\lambda}$:
\begin{equation}
\Gamma_{i,j,\lambda}= \sqrt{F_{i,\lambda} F_{j,\lambda}} \ \Re [ V_{i,j,\lambda} ] + i \sqrt{F_{i,\lambda} F_{j,\lambda}} \ \Im [ V_{i,j,\lambda} ]\,\end{equation}
using $ \sqrt{F_{i,\lambda} F_{j,\lambda}} =  \sqrt{\langle |E_i|^2 \rangle \langle |E_j|^2 \rangle}  $.
All these values are computed in real time for each of the six wavelengths in the K band. In case of split polarization (when the Wollaston is inserted), the calculation is also done independently for both polarizations.

\subsection{Flux estimator}
\label{sec:flux}

The flux $F_i$ is the value extracted after each DIT from Eq~(\ref{eq:P2VM}), summed over the $N_\lambda=6$ spectral channels: 
\begin{equation}
F_i = \sum_{\lambda=1}^{N_\lambda} F_{i,\lambda}\,
,\end{equation}
 where $i$ corresponds to the input beam number.

\subsection{Phase delay estimator}
\label{sec:pd}

The phase delay $\Phi_{i,j}$ is derived from the complex coherent flux, but after a first step to correct for the phase curvature caused by the dispersion:
\begin{equation}
 \Gamma_{i,j,\lambda}'= \Gamma_{i,j,\lambda} \exp \left(iD\left[1-\frac{2.2\mu m}{\lambda}\right]^2\right)
,\end{equation}
which is only a first-order approximation of the dispersion. It is caused both by the atmosphere and by the fibered differential delay lines (FDDL)\footnote{for more information, see the GRAVITY/ESO Final Design Review document {\em VLT-TRE-GRA-15882-6401}.}. 
Therefore, the corrective term $D$ is a time-variable parameter that depends on the position of the star and the position of the FDDLs.
The phase delay is then extracted by coherent addition of the six spectral channels:
\begin{equation}
\Phi_{i,j} = \arg\left(  \sum_{\lambda=1}^{N_\lambda}  \Gamma_{i,j,\lambda}' \right)
.\end{equation}
It is worth noting that the phase delay $\Phi_{i,j} $ is wrapped: it lies between $-\pi$ and $\pi$. No unwrapping effort is made at this stage.

\subsection{Phase variance estimator}
\label{sec:snr}

Computing the S/N of the fringes is essential for the fringe tracker. 
It ensures that the controller does not track on noise. It is also needed for the state machine to know if it has the fringes locked or if it must start looking for the fringes elsewhere.
The S/N is calculated from the variance of the phase delay: $s/n_{i,j}= 1/\sqrt{Var(\Phi_{i,j})}$.
For each DIT, from the photon and background noise, the variance on each pixel is estimated by the relation:
\begin{equation}
Var(q_{i,\lambda})=\sigma_{\rm sky}^2+\gamma ( q_{i,\lambda}- q_{i,\lambda, {\rm sky}} )
,\end{equation}
where $\sigma_{\rm sky}$ and $q_{i,\lambda, {\rm sky}}$ are the noise and flux, respectively, observed during sky observations.  The term $\gamma$ is the detector gain in ADU per detected photon.
The covariance matrix of the real and imaginary part of the $\Gamma$ terms can be obtained from the $\mathsf{P2VM}$:
\begin{equation}
 \mathsf{\Sigma}_\Gamma =
\mathsf{P2VM}_\lambda \cdot \left(
\begin {array}{c}
Var(q_{1,\lambda})\\
\vdots \\
Var(q_{24,\lambda})
\end {array}
\right) \cdot \mathsf{P2VM}_\lambda^\top\,
,\end{equation}
where $^\top$ is the transpose operator. To save processing time, only the diagonal values of the variance matrix are calculated. They correspond to the variance of the real and imaginary parts of $\Gamma_{i,j,\lambda}$. This assumes that the covariance between the real and imaginary parts is negligible (a good assumption for an ABCD with phase shift of $0,\pi/2,\pi, and 3\pi/2$). In the end, for simplicity, we estimated the variance of the phase by the following equation:
\begin{equation}
Var(\Phi_{i,j})=\cfrac{\displaystyle { \sum_\lambda  \langle  Var(\Re\Gamma_{i,j,\lambda} ) + Var(\Im\Gamma_{i,j,\lambda} ) \rangle_{\rm 5 DIT} }  }{ \displaystyle 2 \left| \sum_\lambda  \langle   \Gamma_{i,j,\lambda}' \rangle_{\rm 5 DIT} \right|^2}
\label{eq:varPhi}
,\end{equation}
which is the variance of the amplitude of the coherent flux averaged over five DIT. The five-DIT average is a way to increase the precision of the calculation. However, this is done at the expense of accuracy: the coherent averaging of the complex coherent flux can add a negative bias to the phase variance estimator.

%

\subsection{Group delay estimator}
\label{sec:gd}

The group delay, $\Psi_{i,j}$, is also obtained from the complex coherent flux. Because it consists of a differential measure of the phase as a function of wavelength, this estimator is more noisy than the phase delay \citep{2000SPIE.4006..397L}. To increase its S/N, for each one of the $N_\lambda=6$ spectral channels, the $\Gamma_{i,j,\lambda}$  is first corrected for dispersion, cophased, and averaged over 40 DITs. The result is then used to derive the group delay by multiplying the phasor of consecutive spectral channels:
\begin{eqnarray}
 \Gamma_{i,j,\lambda}''&=& \langle  \Gamma_{i,j,\lambda}' \exp(-i\Phi_{i,j}) \rangle_{\rm 40 DIT} \label{eq:psiavg}\\
\Psi_{i,j} &= &\arg\left(  \sum_{\lambda=1}^{N_\lambda-1} \Gamma_{i,j,\lambda+1}'' \Gamma_{i,j,\lambda}''^* \right)
\label{eq:gd}
.\end{eqnarray}
As for the phase delay, the group delay is estimated modulo $2\pi$. In terms of optical path, however, $\Psi_{i,j}$ corresponds to a value $R$ times smaller than $\Phi_{i,j}$, with $R=23$, the  spectral resolution of the GRAVITY fringe tracker. This is explicit in open-loop operation where the phase delay and group delay are compared for a response to top-hat piezo commands. Because both estimators wrap at $2\pi$, this means that the estimator is valid over a long range equal to 23 times the wavelength. This is the main advantage of the group-delay estimator: to be able to find fringes far away from the central white-light fringe (the fringe of highest contrast). However, because it uses individual spectral channels, the group delay calculated on a single
  DIT would be extremely noisy. Thus the 40 DIT summation is a way to increase the S/N of the group delay, at the cost of losing response time. In the end, the group-delay estimator is a reliable, but slow, estimator of the optical path difference.

\subsection{Closure-phase estimators}
\label{sec:cp}

The closure phases, $\Theta_{i,j,k}$, are calculated on all four triangles from the coherent flux. Before taking the argument, the bispectra are averaged over 300 DIT (corresponding to 330~ms at the fastest 909~Hz sampling rate). Closure phases are estimated from the phase delay:
\begin{equation}
\Theta_{i,j,k}^{\rm PD}=\arg \left( \langle    \sum_{\lambda=1}^{N_\lambda} \Gamma_{i,j,\lambda}'  \sum_{\lambda=1}^{N_\lambda} \Gamma_{j,k,\lambda}'  \sum_{\lambda=1}^{N_\lambda} \Gamma_{i,k,\lambda}'^* \rangle_{\rm 300 DIT}  \right)\,,
\end{equation}
but also from the group delay:
\begin{equation}
\Theta_{i,j,k}^{\rm GD}=\arg \left( \langle    \sum_{\lambda=1}^{N_\lambda-1}  \Gamma_{i,j,\lambda+1}'' \Gamma_{i,j,\lambda}''^* \sum_{\lambda=1}^{N_\lambda-1} \Gamma_{j,k,\lambda+1}'' \Gamma_{j,k,\lambda}''^*  \sum_{\lambda=1}^{N_\lambda-1} \Gamma_{i,k,\lambda+1}''^* \Gamma_{i,k,\lambda}'' \rangle_{\rm 300 DIT}  \right)\,.
\end{equation}
They are  observables modulo $2\pi$, no unwrapping is intended.

       
\section{State spaces, projections, and  state machine}
\label{sec:control1} 

\subsection{OPD-state space}

A main difficulty for the fringe tracker consists of dealing with the different dimensions of the vectors involved. The number of phase observables is six. The number of delay lines is four. Last, the number of degrees of freedom is three. In \citet{2012A&A...541A..81M}, we proposed an $\mathbb{R}^3$ modal-state space orthogonal to the piston space. However, as also mentioned in  \citet{2012A&A...541A..81M}, this modal control has an important drawback: it cannot work in a degraded mode where one or more telescopes are missing. Instead, the GRAVITY controller uses the OPD-state space. In some sense, the implemented fringe tracker is a downgraded version of the state controller proposed in  \citet{2012A&A...541A..81M}.  However, it facilitates managing flux drop-out as well as working with a reduced number of baselines.

\subsection{Reference vectors}
\label{sc:reference}

For the system to work properly, the OPD-state space must be colinear to the piston space. However, because the astronomical object is not necessarily a point source, the closure phases are not necessarily zero. Therefore, the OPD component orthogonal to the piston space must be removed from the measurement. This is done by subtracting a reference position, or set point, which is computed from the closure-phase estimators $\Theta_{i,j,k}^{\rm PD}$ and $\Theta_{i,j,k}^{\rm GD}$. Then, the error terms, that is, the differences between the measured OPD and the set points, are colinear to the piston space.

However, the devil is in the details. There are four closure phases, and only three can be used. The noisiest closure phase is therefore discarded. The three other closure phases are applied on the three edges of the triangle forming the noisiest closure phase. The reference vector is therefore defined as:
\begin{equation}
\mathbf{Ref_\Phi}=
  \left( \begin{array}{c}
  \Theta_{1,2,4}^{\rm PD}\\
\Theta_{1,3,4}^{\rm PD}\\
0\\
\Theta_{2,3,4}^{\rm PD}\\
0\\
0
 \end{array}\right)  
 \text{or}
  \left( \begin{array}{c}
 \Theta_{1,2,3}^{\rm PD}\\
0\\
-\Theta_{1,3,4}^{\rm PD}\\
0\\
-\Theta_{2,3,4}^{\rm PD}\\
0
 \end{array}\right)  
 \text{or}
  \left( \begin{array}{c}
0\\
-\Theta_{1,2,3}^{\rm PD}\\
-\Theta_{1,2,4}^{\rm PD}\\
0\\
0\\
\Theta_{2,3,4}^{\rm PD}
 \end{array}\right)  
 \text{or}
  \left( \begin{array}{c}
 0\\
0\\
0\\
\Theta_{1,2,3}^{\rm PD}\\
\Theta_{1,2,4}^{\rm PD}\\
\Theta_{1,3,4}^{\rm PD}
 \end{array}\right)  
 \label{eq:RefPD}
,\end{equation}
depending on which triangle has the lowest S/N, from left to right, the 123, 124, 134, or 234 triangle. The reference values for the group delay are calculated similarly:
\begin{equation}
\mathbf{Ref_\Psi}=
  \left( \begin{array}{c}
  \Theta_{1,2,4}^{\rm GD}\\
\Theta_{1,3,4}^{\rm GD}\\
0\\
\Theta_{2,3,4}^{\rm GD}\\
0\\
0
 \end{array}\right)  
 \text{or}
  \left( \begin{array}{c}
 \Theta_{1,2,3}^{\rm GD}\\
0\\
-\Theta_{1,3,4}^{\rm GD}\\
0\\
-\Theta_{2,3,4}^{\rm GD}\\
0
 \end{array}\right)  
 \text{or}
  \left( \begin{array}{c}
0\\
-\Theta_{1,2,3}^{\rm GD}\\
-\Theta_{1,2,4}^{\rm GD}\\
0\\
0\\
\Theta_{2,3,4}^{\rm GD}
 \end{array}\right)  
 \text{or}
  \left( \begin{array}{c}
 0\\
0\\
0\\
\Theta_{1,2,3}^{\rm GD}\\
\Theta_{1,2,4}^{\rm GD}\\
\Theta_{1,3,4}^{\rm GD}
 \end{array}\right)  
 \label{eq:RefGD}
.\end{equation}
 The closure-phase changes as a function of time, causing the reference position to adapt to any change in the phase closures. The closure phase is smoothed over a long enough time (300\,DIT) to avoid adding additional noise.
 However, the choice of which triangle is the noisiest is made only once between each scientific frame to avoid sharp jumps in the reference vector over the integration time of the science detector. 

This reference scheme works most of the time. However, problems arise in two specific instances. First, when the object is so hgihgly resolved that the used closure phases contain a   baseline with zero visibility. If that happens, an undefined reference value is applied to a perfectly sane baseline and the system can diverge. Second, if the fringe tracker is tracking on two unconnected baselines (e.g., between telescopes 1-2 and 3-4), the closure phases are undefined, and using their values would mean losing one of the two locked baselines. To resolve these two problems, the closure phases are modified as follows:
if  one of the baselines of any of $ij$, $jk,$ or $ik$ have an S/N below the value $s/n_{\rm threshold}^{\rm GD}$, then  $\Theta_{i,j,k}^{\rm PD}$ used in Eq.~(\ref{eq:RefPD}) is a fixed value, and $\Theta_{i,j,k}^{\rm GD}=0$ is used in Eq.~(\ref{eq:RefGD}). The difference in treatment between the group and phase delay arises because the default group-delay tracking shall be zero, while the default phase-delay tracking can be any constant value.

\subsection{Transfer matrices}
    
After the reference values are subtracted from the OPD, we can freely project the data in piston-state space as well as back to the OPD-state space.
Hereafter, we use the same nomenclature as in  \citet{2012A&A...541A..81M}. $\mathbf P$ corresponds to the four-dimension piston-state vector, while $\mathbf {OPD}$ corresponds to the six-dimension OPD-state vectors. The matrix $\mathsf M$ corresponds to the conversion between piston and optical path difference:
\begin{equation}
\mathbf{OPD}=\mathsf M\,\mathbf P,\label{eq:OPDMP}\end{equation}
where
\begin{equation}\mathsf
M=\begin{pmatrix}-1&-1&-1&0&0&0\\1&0&0&-1&-1&0\\0&1&0&1&0&-1\\0&0&1&0&1&1
\end{pmatrix}^\top.\label{eq:Mmatrix}
\end{equation}
The conversion $\mathbf {OPD}\rightarrow \mathbf{P}$ is ill-constrained, however: the rank of matrix $\mathsf M$ is 3, not 4. This is because the global piston cannot be obtained from the differences in the optical path. Nevertheless, we can  define a pseudo-inverse matrix:
\begin{equation}
\mathsf M^\dag=\frac14\begin{pmatrix} -1&-1&-1&0&0&0\\
1&0&0&-1&-1&0\\
0&1&0&1&0&-1\\
0&0&1&0&1&1
\end{pmatrix}.
\end{equation}
where $^\dag$ denotes the pseudo-inverse operator. 

\subsection{Thresholds and S/N management}
\label{sec:I}

   \begin{figure}
   \centering
   \includegraphics[width=0.5\textwidth]{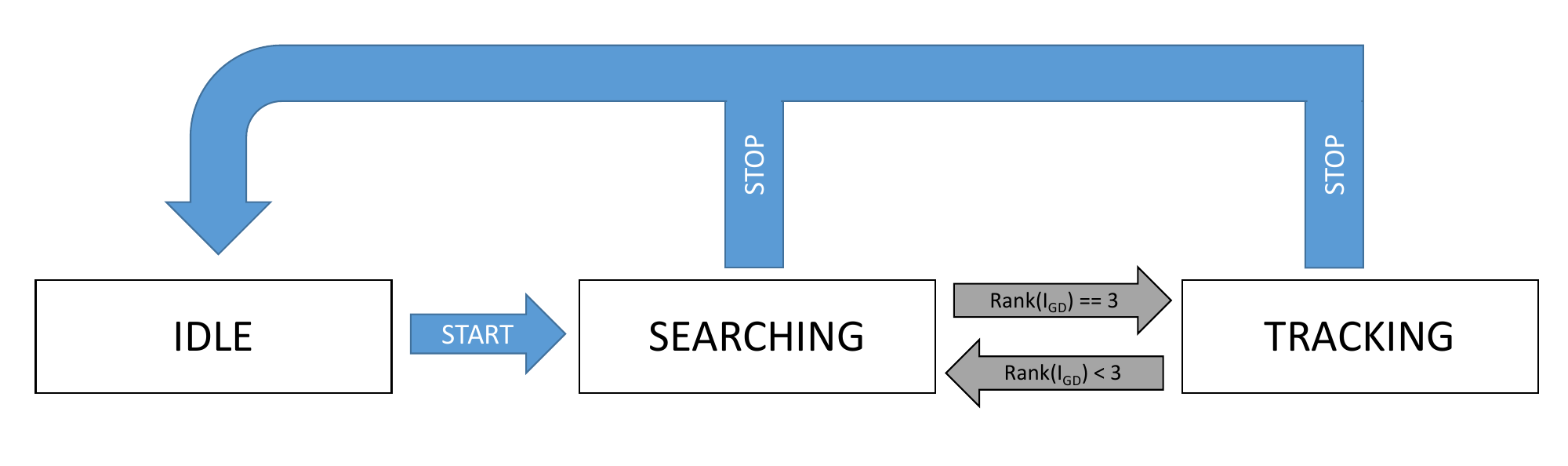}
   \caption{State machine of the fringe tracker. There are only three states: IDLE, SEARCHING, and TRACKING. The blue transitions are commands from the operator. The gray transitions are automatic decisions. Switching from searching to tracking only depends on the rank of the $\mathbb I_{\rm GD}^4$ matrix. If GRAVITY is running in a degraded mode, the TRACKING transition can happen for a rank of 2 or even 1.}
              \label{fig:stateMachine}%
    \end{figure}
    
The system uses two distinct thresholds. A first threshold, 
$s/n_{\rm threshold}^{\rm GD}$, disables the baselines where the S/N is too low to make the baseline useful. This is the case when the fringes are not yet found, 
when the fringes are suddenly lost, or when the astronomical target is so highly resolved that the spatial coherence is close to zero. To detect these events, the
  $s/n_{\rm threshold}^{\rm GD}$ is compared to a moving average of the phase-delay variance estimator:  $\langle Var(\Phi_{i,j}) \rangle_{\rm 40 DIT}$. This group-delay threshold must be adapted to the target: it mustbe high enough to ensure that the fringe tracker does not track on side-lobes, but low enough to detect the fringes.

A second threshold, $s/n_{\rm threshold}^{\rm PD}$, is used solely for the phase-tracking controller. It disables the tracking on a given telescope in case of rapid S/N drop-off. The main target of this threshold is to be able to catch a drop in flux injection (caused by external tip-tilt), with the expectation that the S/N will increase soon hereafter. Typically, $s/n_{\rm threshold}^{\rm PD}=1.5$ and $s/n_{\rm threshold}^{\rm PD} < s/n_{\rm threshold}^{\rm GD}$.

Hence, we defined two matrices $\mathbb I$ that convert OPD space error into a space of the same dimension colinear to the OPD space. They read
\begin{eqnarray}
\mathbb I_{\rm GD}^6&=&\mathsf M (\mathsf M^\top\mathsf W\,\mathsf M)^{\dag_{\rm GD}}\,\mathsf M^\top\mathsf W \\
\mathbb I_{\rm PD}^6&=&\mathsf M (\mathsf M^\top\mathsf W\,\mathsf M)^{\dag_{\rm PD}}\,\mathsf M^\top\mathsf W 
,\end{eqnarray}
where the only difference is on the pseudo-inverse operator $^\dag$.
In both equations, the $6\times6$ weighting matrix $\mathsf W$  distributes the weights among the different OPDs:
\begin{equation}
\mathsf W=\mathrm{diag} 
\left( 
\begin{array}{cccccc}
w_{1,2}& w_{1,3} & w_{1,4} & w_{2,3} & w_{2,4} & w_{3,4} \\
\end{array}
\right)
,\end{equation}
where
 \begin{equation}
w_{i,j}= \begin{cases}
0 &\text{if $ 1/{\langle Var(\Phi_{i,j}) \rangle_{\rm 40 DIT}} < |s/n_{\rm threshold}^{\rm GD}|^2  $} \\
1/Var(\Phi_{i,j})   &\text{otherwise}
\end{cases}
.\end{equation}
This step is important to remove the risk of tracking on noise: if the variance of the phase reaches this threshold, the pseudo-inverse discards that baseline in its calculation. The pseudo-inversion is done using a SVD:
\begin{equation}
\mathsf M^\top\mathsf W\,\mathsf M= \mathsf U \mathsf S \mathsf V^\top 
.\end{equation}
The idea behind this decomposition is that $ \mathsf  U$ and $ \mathsf V$ are two invertible orthonormal matrices (the left-singular and right-singular eigenvectors, respectively): $\mathsf I=\mathsf U \mathsf U^\top$ and $\mathsf I=\mathsf V \mathsf V^\top$ , where $\mathsf I$ is the identity matrix.
  $\mathsf S$ is a square diagonal matrix where the values on the diagonal correspond to the square root of the eigenvalues:  $\mathsf S= \mathrm{diag} ( s_1, s_2, s_3, 0)$. For a four-telescope operation, three eigenvalues are non-zero. The number of non-zero eigenvalues decreases when the system cannot track all telescopes.

The pseudo-inverse of matrix $\mathsf S$ is calculated differently for the group-delay and phase-delay control loop. For the group-delay control loop, we have
  $\mathsf S^{\dag_{\rm GD}} = \mathrm{diag} ( s_1^{\dag_{\rm GD}} , s_2^{\dag_{\rm GD}} , s_3^{\dag_{\rm GD}} , 0),$ where
 \begin{equation}
s_i^{\dag_{\rm GD}} = \begin{cases}
1/s_i  &\text{if $ s_i > 0$  } \\
0 &\text{otherwise}
\end{cases}
.\end{equation}
For the phase-delay control loop, we have
  $\mathsf S^{\dag_{\rm PD}} = \mathrm{diag} ( s_1^{\dag_{\rm PD}} , s_2^{\dag_{\rm PD}} , s_3^{\dag_{\rm PD}} , 0),$ where
 \begin{equation}
s_i^{\dag_{\rm PD}} = \begin{cases}
1/s_i  &\text{if $ s_i > |s/n_{\rm threshold}^{\rm PD}|^2$  } \\
s_i / |s/n_{\rm threshold}^{\rm PD}|^4  &\text{otherwise}
\end{cases}
\label{eq:sinvPD}
.\end{equation}

As a result, the two matrices $\mathbb I$ are calculated for each DIT from a new SVD and the equations
\begin{eqnarray}
\mathbb I_{\rm GD}^6&=&\mathsf M  \mathsf V   \mathsf S^{\dag_{\rm GD}}    \mathsf U ^\top  \mathsf M^\top\mathsf W \\
\mathbb I_{\rm PD}^6&=&\mathsf M  \mathsf V    \mathsf S^{\dag_{\rm PD}}    \mathsf U ^\top  \mathsf M^\top\mathsf W \label{eq:sinvPD2}
,\end{eqnarray}
where the difference between the two is that the eigenvalues in $\mathbb I_{\rm PD}^6$ are weighted down when they are below a value equal to $|s/n_{\rm threshold}^{\rm PD}|^2$.

\subsection{State machine}
\label{sec:fringeSearch}

The rank of matrix $\mathbb I_{\rm GD}^6$ (the number of non-zero eigenvalues) drives the decision-making of the state machine.
 For a four-telescope operation, a rank of 3 means that the position of the delay lines on all the telescopes is constrained. When only three telescopes are tracked, the rank is 2. The rank is 1 for two linked telescopes.

The state machine (Fig.~\ref{fig:stateMachine}) therefore has only three states: IDLE, SEARCHING, and TRACKING. 
When the operator starts the fringe tracker, it switches to the state SEARCHING. As soon as the rank of the $\mathbb I_{\rm GD}^6$ matrix is 3, the fringe tracker transitions to state TRACKING. When the rank of the matrix decreases and remains low for a period of 1\,s or more, then the system automatically switches back to SEARCHING mode. In both states, the group-delay and phase-delay controllers are running. This means that whether in SEARCHING or TRACKING state, the system still tracks the fringes on the baselines with sufficient S/N.


\section{Group delay tracking}
\label{sec:control2}

\subsection{Group-delay block diagram}

\begin{figure}
   \centering
   \includegraphics[width=0.5\textwidth]{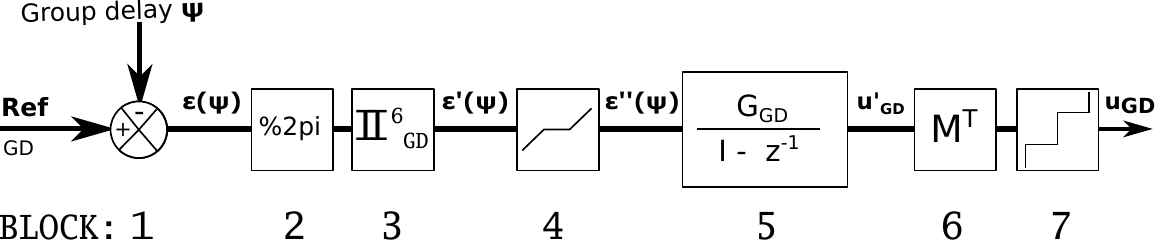}
   \caption{Block diagram of the group-delay controller. The controller is centered on an integrator (block  5), but also includes several features to work in conjunction with the phase-delay controller.}
              \label{fig:GDblocks}%
    \end{figure}

Fig.~\ref{fig:control} represents the block diagram of the full control architecture. Within this design,
the prevalent control loop is the group-delay loop (in blue). It has to be very reliable: the signal on the feedback signal $\mathbf{\Psi}$ is therefore enhanced by averaging over 40 DITs in Eq.~(\ref{eq:psiavg}). The control algorithm of the group delay is presented in Fig.~\ref{fig:GDblocks}.  It generates a control signal $\vec u_{\rm GD}$ whose unit is the radian of the phase delay.
It is made of seven distinct blocks that are described below.

The measured group delay is a vector defined as
\begin{equation}
\vec \Psi_n=\left( 
\begin{array}{cccccc}
\Psi_{1,2} & \Psi_{1,3} & \Psi_{1,4} & \Psi_{2,3} & \Psi_{2,4} & \Psi_{3,4}  \\
\end{array}
\right)_n\,.
\end{equation}
The first block is a comparator that extracts the error between a reference vector ($\mathbf{Ref_\Psi}$) and the measured group delay.  The logic would be to have all six setpoints equal to zero to track on the white-light fringes. However, as explained in Sec.~\ref{sc:reference}, this is not possible in the presence of non-zero group-delay closure phase $\Theta^{\rm GD}$. The use of a setpoint vector as defined by Eq.~(\ref{eq:RefGD}) therefore ensures that all baselines track the fringes with the highest contrast that do not contradict each other:
\begin{equation}
\vec \varepsilon _{\vec \Psi,n}  = \mathbf{Ref_\Psi} - \vec \Psi_{n} \,.
\end{equation}

The second block is there because the phase measurement is only known modulo $2\pi$. Because $\vec \varepsilon _{\vec \Psi,n}$ can have any value, this block adds or subtracts an integer number of $2\pi$ to ensure that the error is between $-\pi$ and $\pi$.

The third block uses the $\mathbb I_{\rm GD}^6$ matrix to weight the errors between the different baselines. Following that matrix, the OPD error vector is now strictly, in a mathematical sense, colinear to the piston space. Moreover, the error on any baseline with no fringes is either estimated from other baselines or set to zero. After the third block, the error group-delay vector is now
\begin{equation}
\vec {\varepsilon '}_{\vec \Psi,n}  = \mathbb I_{\rm GD}^6 ([ \mathbf{Ref_\Psi} - \vec \Psi_{n} ]\% 2\pi) \,,
\end{equation}
where the percent sign corresponds to the modulo function.

The fourth block is a threshold function that quenches the gain of the control loop if the absolute value of $\vec {\varepsilon'}_{\vec \Psi,n}$ is lower than $2\pi/R$. This value corresponds to an OPD of one wavelength, meaning the group-delay controller cannot converge on an accuracy better than $2.2\,\mu$m. This is necessary to let the phase-delay controller track within a fringe:
\begin{equation}
\vec {\varepsilon '' }_{\vec \Psi,n} = \begin{cases}
\vec {\varepsilon '}_{\vec \Psi,n} - \pi/R  &\text{if \ $\vec {\varepsilon'}_{\vec \Psi,n} > \pi/R $  } \\
\vec {\varepsilon '}_{\vec \Psi,n} + \pi/R  &\text{if \ $\vec {\varepsilon'}_{\vec \Psi,n} < -\pi/R $  } \\
0 &\text{otherwise.}
\end{cases}
\end{equation}

The fifth block is the integrator. In the time domain, it writes\begin{equation}
\vec {u'}_{\mathrm{GD},n}=\vec {u'}_{\mathrm{GD},n-1}+\mathsf G_{\rm GD}\, \vec {\varepsilon ''}_{\vec \Psi,n} 
.\end{equation}
The same controller gain $\mathsf G_{\rm GD}$ is applied to all baselines. It minimizes the closed-loop response time and maximizes robustness.
The open-loop transfer function is the same for each baseline. It is mostly a pure delay caused by the moving average of 40 DITs, as stated by Eq.~(\ref{eq:psiavg}) in Sect.~\ref{sec:gd}. The $-3$\,dB cutoff frequency therefore depends on the sampling rate of the fringe tracker. It is 13, 4.5, and 1.5\,Hz for sampling rates of 909, 303, and 97\,Hz, respectively. 

The sixth block is the matrix $\mathsf M^\dag$ to transpose the control signal from OPD space to piston space.

 The last block is a quantization function. Practically, it means that the group-delay controller, when it detected a group-delay error larger than a fringe, only makes $2\pi$ phase-delay jumps until it comes to the reference fringe, without disturbing the long-term phase measurement.  The command issued from the group-delay controller is therefore for each piezo-actuator the closest value that is a multiple of $2\pi$. Hence
\begin{equation}
\vec u_{\mathrm{GD},n}= \mathrm{round}_{ \{ 2\pi \} } \left( \mathsf M^\dag \,   \vec {u'}_{\mathrm{GD},n} \right)\,.
\end{equation}
When $\mathsf M^\dag \,   \vec {u'}_{\mathrm{GD},n}$ is between $-\pi$ and $\pi$, the control signal of the group-delay controller is constant, and the phase delay can work without  interferences caused by the group-delay loop.

We note that the final control signal (as shown in Fig.~\ref{fig:control}) also includes  the modulation function, the fringe-search function, and the phase-delay control signal:
\begin{equation}
\vec u = \vec u_{\mathrm{GD},n} + \vec u_{\mathrm{modulation}}  + \vec u_{\mathrm{search},n} +\vec u_{\mathrm{PD},n} \,. 
\end{equation}

\subsection{Modulation function and $2\pi$ phase jumps}
\label{sec:fringe}

   \begin{figure}
   \centering
   \includegraphics[width=0.5\textwidth]{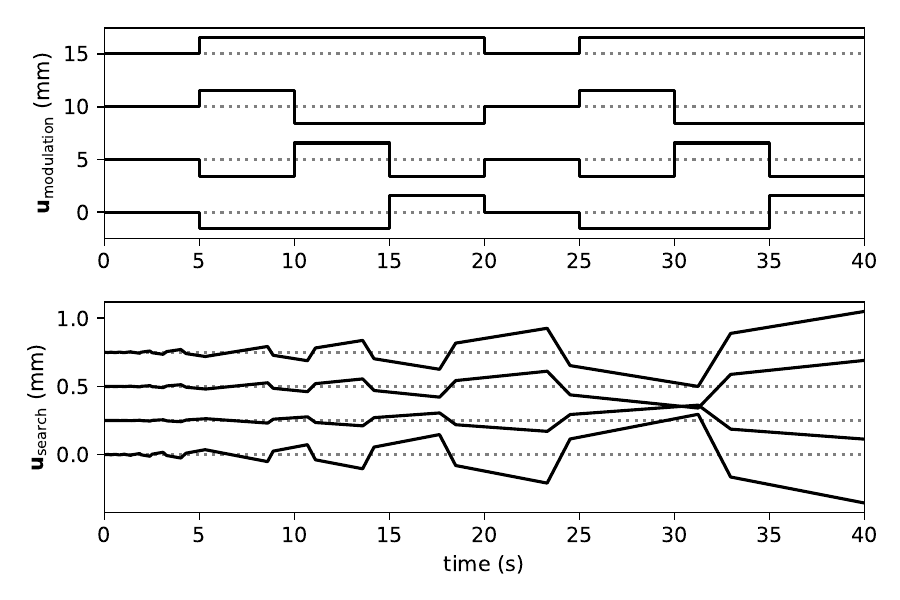}
   \caption{ \textbf{Upper panel:} Signal from the modulation block, in radians (the Volts-to-radians gain is not accounted for here). At each science exposure (here 5 seconds), the  $\vec u_{\mathrm{modulation}}$ values change to make $\pi$ offsets between the baselines. The signal repeats every four science exposures so that each baseline is observed with as many  $+\pi/2$ as $-\pi/2$ offsets. \textbf{Lower panel:} Modified sawtooth signal to search for fringes when the fringe tracker is in SEARCHING state. Here the research is made on all telescopes, meaning that the rank of matrix $\mathbb I_{\rm GD}^6$ is 0.}
              \label{fig:ucom}%
    \end{figure}
    
The rounding of the group-delay command ensures that the same phase is always tracked even though jumps between fringes are required when the white-light fringe is searched for. However, this means that the science detector always records the fringes at the same phase delay, and when the sky is not well subtracted, for example, this can bias the visibility. This can  be explained in the case of a perfect ABCD beam combiner. Assuming the recorded flux on each one of the four pixels is $q_A$, $q_B$, $q_C$ , and $q_D$, the $\mathsf {P2VM}$ matrix calculates the raw complex visibility in this way:
\begin{equation}
V=q_A-q_C + i (q_B- q_D )\,.
\end{equation}
When the sky removal on $q_A$ is not perfect, we have an additional flux that biases the measurement: $\hat q_A=q_A+ \varepsilon_A$. To remove this error term, the solution is to record the fringes with $\pi$ offsets:
\begin{eqnarray}
\hat V&=&q_A+\varepsilon_A-q_C + i (q_B- q_D )\\
\hat V_\pi&=&q_C+\varepsilon_A-q_A + i (q_D- q_B)
,\end{eqnarray}
giving\begin{equation}
\frac{\hat V-\hat V_\pi}{2} = \frac{2q_A-2q_C+ i (2q_B- 2q_D)}{2}=V \,.
\end{equation}
This  temporal $\pi$ modulation is added to the group-delay command (Fig.~\ref{fig:ucom}). It is not seen by the group-delay control loop because of the quenching block (block 4 in Fig.~\ref{fig:GDblocks}). This command is synchronized with the reset of the science detector, and it is sequentially:
\begin{equation}
 \vec u_{\mathrm{modulation}}  = 
  \left( \begin{array}{c}
 0 \\ 0 \\ 0 \\ 0 \end{array}\right)  \text{or}
  \left( \begin{array}{c}
  \pi/2 \\  \pi/2 \\  - \pi/2 \\   - \pi /2 \end{array}\right)  \text{or}
  \left( \begin{array}{c}
  \pi/2 \\  - \pi/2 \\   \pi/2 \\   - \pi /2 \end{array}\right)  \text{or}
  \left( \begin{array}{c}
  \pi/2 \\  - \pi/2 \\  - \pi/2 \\    \pi /2 \end{array}\right)  \,.
  \label{eq:seq}
\end{equation}
Only with a minimum of 4 DIT with each of the $\vec u_{\mathrm{modulation}}$ above can we have, on each baseline, as many exposures with 0 and $\pi$ offsets. Therefore
the number of science DITs within a GRAVITY exposure is recommended to be a multiple of 4.

\subsection{Fringe search}

   \begin{figure}
   \centering
   \includegraphics[width=0.5\textwidth]{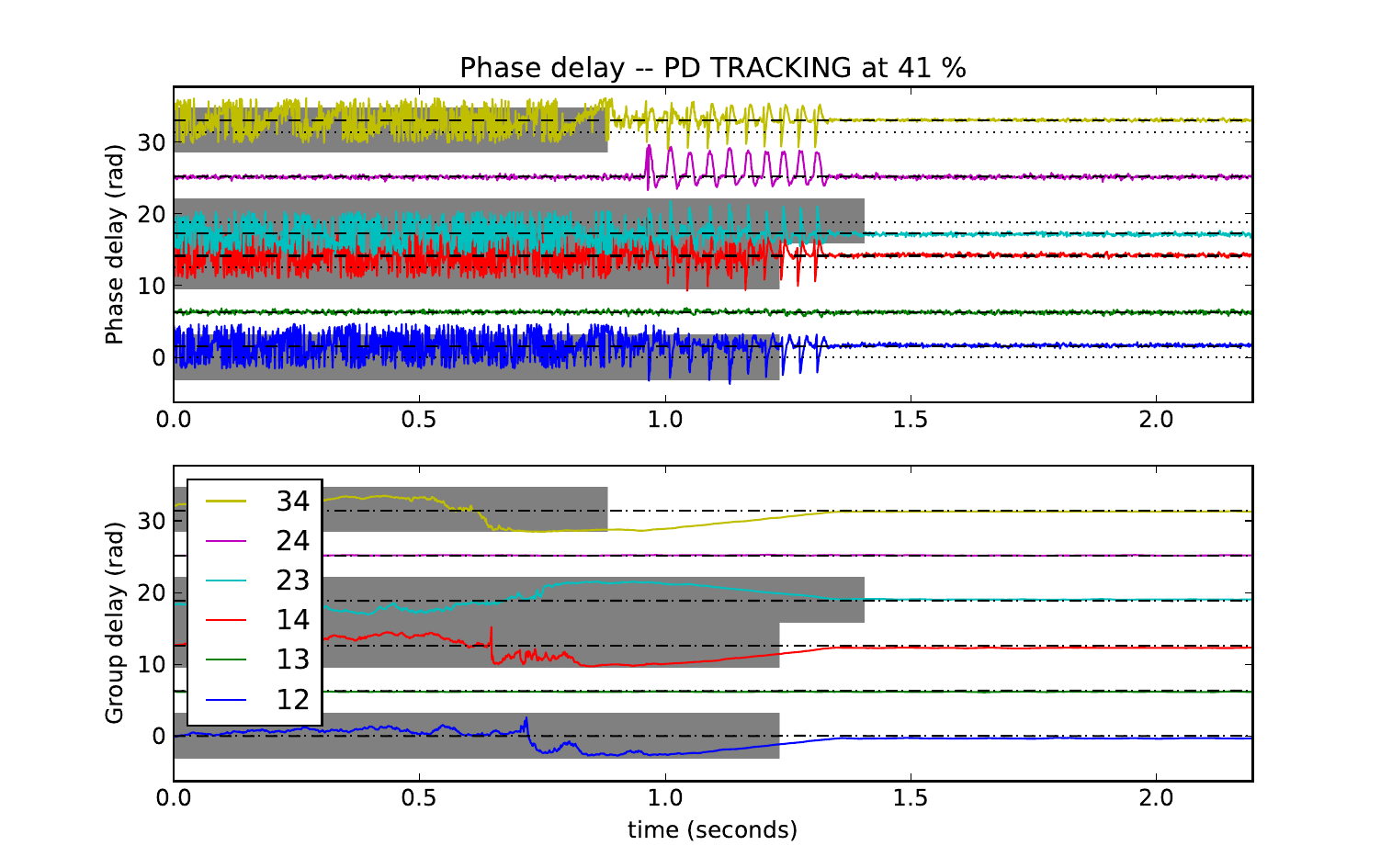}
   \caption{
   Phase $\Phi_{i,j}$ (upper panel) and group $\Psi_{i,j}$ (lower panel) delays. All values are phases modulo $2\pi$. The dashed lines correspond to the piezo command. 
   From bottom to top, the phases correspond to baselines $i,j$ equal to 12, 13, 14, 23, 24, and 34.
   At the beginning of this recording, the system only has fringes on two baselines (green and purple). At $t=0.8\,$s, the system found fringes on the yellow baseline, and later on all the other baselines. Gray areas correspond to the baselines whose S/N is below the value $s/n_{\rm threshold}^{\rm GD}$, meaning that no tracking is performed. After $t=1.4\,$s, the system nominally tracks on all baselines.}
              \label{fig:catch}%
    \end{figure}

The search is made through a modified  sawtooth function of increasing amplitude ($u_\mathrm{saw} $). This function is started when the fringe tracker enters SEARCHING state. It is disabled when the system transition to TRACKING state. This is the only difference between the two modes.  The same sawtooth function is generated for each piezo-actuator, but with different scaling factors: $-2.75$, $-1.75$, $1.25,$ and $3.25$ for the first, second, third, and fourth beam, respectively. The four commands are then multiplied by the kernel of the $\mathbb I_{\rm GD}^4$ matrix. This matrix is computed as
\begin{equation}
\mathbb I_{\rm GD}^4 = \mathsf M \mathbb I_{\rm GD}^6 \mathsf M^\dag
\end{equation}
giving the following signal:
\begin{equation}
 \vec u_{\mathrm{search},n}  = (\mathsf I^4-\mathbb I_{\rm GD}^4) 
 \left( \begin{array}{cccc}
-2.75 &  -1.75& 1.25 & 3.25  \end{array}
\right)^\top u_\mathrm{saw} 
,\end{equation}
where $\mathsf I^4$ is the $4\times4$ identity matrix. This ensures that the baselines with sufficient fringe signal are tracking and are not modulated by the sawtooth function. When no more fringes are found, the kernel is the identity, and each baseline is modulated. This gives the trajectory shown in Fig~\ref{fig:ucom}.

An example of fringe research and acquisition is presented in Fig.~\ref{fig:catch}.
 In this example, at $t=0$, the fringes are found and tracked on two baselines: the baseline between telescopes 1 and 3, and the baseline between telescopes 2 and 4. The group delay for the two fringes is zero, and the rank of the $\mathbb I_{\rm GD}^4$ matrix is 2. The fringe tracker is therefore in  SEARCHING state. At $t=0.8\,$s, the fringes are found on baseline 34. The rank of the $\mathbb I_{\rm GD}^4$ matrix increases, and the system switches to TRACKING state. The group delays of baselines 12, 14, 23, and 34 are brought to zero. Signals are later found on all six baselines and the system reaches a nominal tracking state at $t=0\,$s.

       
\section{Phase-delay tracking}
\label{sec:control3} 

\subsection{Principle}

The GRAVITY phase control loop uses both piston-space and OPD-space state vectors. This is the easiest way to  properly handle  both piezo-actuators and the atmosphere dynamics \citep{2008SPIE.7015E..1FC}.
The vibrations and atmospheric perturbations are represented by six OPD-space state vectors labeled  together $\vec x_V$. Each vector corresponds to a baseline. The piezo-actuators are characterized by four piston-space state vectors $\vec x_P$, one for each delay line. The phase delay, $\vec \Phi_n$, is a vector  of  measured phases:
\begin{equation}
\vec \Phi_n=\left( 
\begin{array}{cccccc}
\Phi_{1,2} & \Phi_{1,3} & \Phi_{1,4} & \Phi_{2,3} & \Phi_{2,4} & \Phi_{3,4}  \\
\end{array}
\right)_n
,\end{equation}
which results from a linear combination of the state vectors $\vec x_{V,n}$ and $\vec x_{P,n}$ at a time $n$.

\begin{figure}
   \centering
   \includegraphics[width=0.5\textwidth]{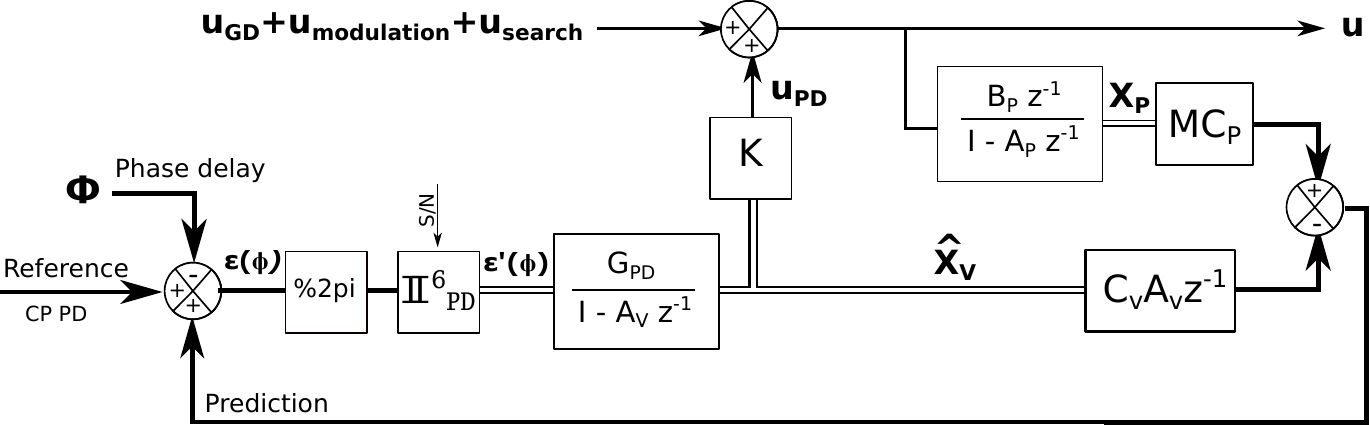}
   \caption{Block diagram of the phase-delay controller. The two control signals $\vec u_{\rm GD}$ and $\vec u_{\rm PD}$ are summed before they are applied to the actuators. In this control scheme, the state vectors $\hat{ \vec X}_V$  are regulated. The control signal $\vec u_{\rm PD}$ is then issued from a matrix multiplication of the state vectors: $\vec u_{\rm PD}=\mathsf K \hat{ \vec X}_V$.}
              \label{fig:PDblocks}%
    \end{figure}

The real-time algorithm of the fringe tracker follows a sequence:
\begin{enumerate}
\item it predicts the future state of the system from previous state using the equation of state (Sec.~\ref{sec:eqState}),
\item it uses Kalman filtering to update the state vectors  (Sec.~\ref{sec:kalFilter}), and
\item it uses the system state to command the piezo-actuators to correct for vibrations and atmospheric effects.  (Sec.~\ref{sec:kalCommand}).
\end{enumerate}
The phase-delay controller is summarized by the block diagram presented in Fig.~\ref{fig:PDblocks}.

\subsection{Equations of state}
\label{sec:eqState}

The equations of state are
\begin{eqnarray}
\vec x_{V,n+1}  &=&{\mathsf A_V} \cdot \vec x_{V,n} + {\mathsf B_V} \cdot \vec v_n \label{eq:state} \\
\vec x_{P,n+1}  &=&{\mathsf A_P} \cdot \vec x_{P,n} + {\mathsf B_P} \cdot  \vec u_n \,,\end{eqnarray}
where each matrix is three-dimensional. Because we assumed uncorrelated baselines, only the diagonals are populated:
\begin{eqnarray}
\mathsf A_V&=&\mathrm{diag} \left(\begin{array}{cccccc}
\mathsf A_V^{1,2} & \mathsf A_V^{1,3} & \mathsf A_V^{1,4}& \mathsf A_V^{2,3} & \mathsf A_V^{2,4}&  \mathsf A_V^{3,4} \end{array}
\right)\\
\mathsf B_V&=&\mathrm{diag} \left( \begin{array}{cccccc}
\mathsf B_V^{1,2} & \mathsf B_V^{1,3} & \mathsf B_V^{1,4}& \mathsf B_V^{2,3} & \mathsf B_V^{2,4}&  \mathsf B_V^{3,4} \end{array}
\right)\\
\mathsf A_P&=&\mathrm{diag} \left( \begin{array}{cccc}
\mathsf A_P^{1} & \mathsf A_P^{2} & \mathsf A_P^{3}& \mathsf A_P^{4}  \end{array}
\right)\\
\mathsf B_P&=&\mathrm{diag} \left( \begin{array}{cccc}
\mathsf B_P^{1} & \mathsf B_P^{2} & \mathsf B_P^{3}& \mathsf B_P^{4} \end{array}
\right)
.\end{eqnarray}
Here the upper index corresponds to the telescope or baseline numbers. In the block diagram of Fig.~\ref{fig:control}, the equations of state are written in the frequency domain, but the transfer function writes
\begin{eqnarray}
\mathcal Z \{\vec x_{V,n}\}  &=&\frac{{\mathsf B_V}  \, z^{-1}}{1-{\mathsf A_V}  \, z^{-1} } \mathcal Z \{\vec v_{n}\} \label{eq:Zv}\\
\mathcal Z \{\vec x_{P,n}\}  &=&\frac{{\mathsf B_P}  \, z^{-1}}{1-{\mathsf A_P}  \, z^{-1} } \mathcal Z \{\vec u_{n}\} \,.
\end{eqnarray}

The evolution of the system is driven on one hand by white noise $\vec v_n$, and on the other hand by a user-controlled voltage applied to the piezo-actuators $\vec u_n$. When $\vec v_n$ is white noise, $ \vec x_{V,n}$ has a colored noise, as highlighted by Eq.~(\ref{eq:Zv}). The problem is similar for AO systems, where it has already been mentioned and corrected for  \citep{2010JOSAA..27A.223P}.

In  \citet{2012A&A...541A..81M} and \citet{2014A&A...569A...2C}, we have shown that using several autoregressive (AR) models of order 2 in parallel was effective to correct for both vibration frequencies and atmospheric turbulence. However, practically, two issues made that implementation difficult: i) the determination of the vibration peaks for low S/N data, and ii) the need to change the space model when vibrations appear or disappear. Both problems can be technically resolved, but to ensure maximum robustness, we preferred a fixed well-defined space model. The idea is that the state  $x_V$  do change with time, but the space-state model does not.
We therefore used an autoregressive model of order 30. This means that the state-space model corresponds to the 30 last values of the phase delay:
\begin{equation}
\mathsf A_V^{i,j} = 
\left(
\begin{array}{cccccc}
v_1^{i,j}  & v_2^{i,j}  & v_3^{i,j}  & \cdots & v_{29}^{i,j}  & v_{30}^{i,j}  \\
1 & 0 & 0 & \cdots &  0 & 0\\
0 & 1 & 0 & \cdots &  0 & 0\\
0 & 0 & 1 & \cdots &  0 & 0\\
\cdots & \cdots & \cdots & \cdots &  \cdots & \cdots \\
0 & 0 & 0 & \cdots &  1 & 0
\end{array}
\right)
\end{equation}
and
\begin{equation}
\mathsf B_V^{i,j} =\left( 
\begin{array}{cccccc}
1 &0 & 0 &\cdots& 0 & 0 \\
\end{array}
\right)^\top
.\end{equation}
The value 30 allows for complexity in the vibrational pattern while characterizing the perturbations with a sufficiently low degree of freedom. 
It was chosen in light of \citet{2012SPIE.8447E..0ZK}: they showed that it is a good compromise with respect to other state-space models to correct atmosphere and vibration for the tip-tilt of AO systems.
The transfer matrix $A_V^{i,j}$ is determined every 5 seconds on a workstation that is connected on the reflective memory network. 

\begin{table}
\caption{Open-loop transfer function parameters of the fringe tracker running at 909~Hz}            
\label{tb:gain}      
\centering                          
\begin{tabular}{c c c c c}        
\hline\hline                 
\noalign{\smallskip}
          (rad/Volts)   & $ i= 1$ &  $i=2$&  $i= 3$ & $ i= 4 $\\
\noalign{\smallskip}
\hline                        
\noalign{\smallskip}
a$_1^i$  & 0.16  &-0.03 & -0.22  &-0.1 \\
a$_2^i$   &-0.15   &0.09  & 0.11  & 0.12\\
a$_3^i$   &  6.52  &4.08   &3.56   &4.41\\
a$_4^i$  & 9.61   &9.01  & 7.12  &11.14\\
a$_5^i$   &  1.31  & 4.72  & 7.05 &  1.85\\
\noalign{\smallskip}
\hline                                   
\noalign{\smallskip}
$G^i_{\rm piezo}$  &  17.47  & 17.89  & 17.63 &  17.43\\
\noalign{\smallskip}
\hline                                   
\end{tabular}
\end{table}

\begin{figure}
   \centering
   \includegraphics[width=0.5\textwidth]{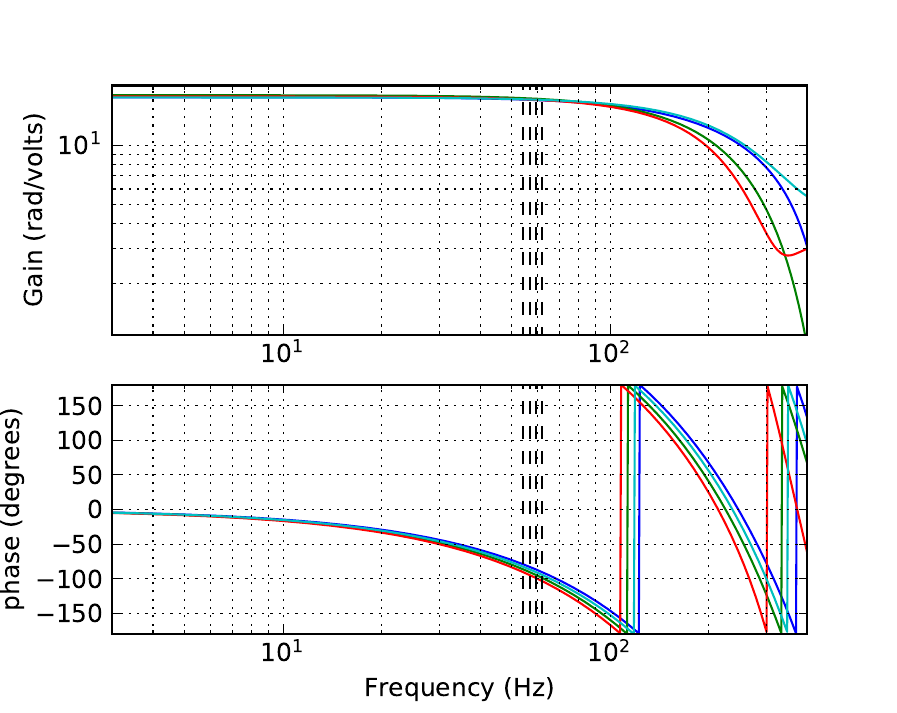}
   \caption{Bode plot of the frequency response of the four pistons in open loop (including pure delay caused by integration and computation). The -90 degree cutoff frequencies are 61Hz, 56 Hz, 54 Hz, and 59Hz. The cutoff frequency is dominated by the pure delay caused by the control loop system.}
              \label{fig:bode}%
    \end{figure}

The transmission matrix $A_P^i $ is also an autoregressive model, but of order 5. It includes the response function of the full system (image integration time, processing time, inertia of the piezo mirror, etc.), from setting a control voltage $\vec u_n$ to measuring a phase delay $\vec \Phi_n$.
 It is measured monthly by injecting a top-hat signal into the system. The matrix for the AR5 model verifies 
\begin{equation}
\mathsf A_P^{i}  = 
\left(
\begin{array}{ccccc}
a_1^i & a_2^i  & a_3^i  & a_4^i  & a_5^i  \\
1 & 0 & 0 & 0 & 0\\
0 & 1 & 0 & 0 & 0\\
0 & 0 & 1 & 0 & 0\\
0 & 0 & 0 & 1 & 0
\end{array}
\right)
\end{equation}
and
\begin{equation}
\mathsf B_P^i =\left( 
\begin{array}{ccccc}
1 &0 & 0 & 0 & 0 \\
\end{array}
\right)^\top
,\end{equation}
where the $a_0^i$, $a_1^i$, $a_2^i$, $a_3^i$ , and $a_4^i$ values, obtained experimentally on the fringe tracker are presented without normalization in Table~\ref{tb:gain}. 
The same values can be represented in a Bode plot, see Fig.~\ref{fig:bode}. The cutoff frequencies, calculated at a phase of -90 degrees, are around 60~Hz. This is considerably below the bandwidth of the piezo-actuator measured by \cite{2010SPIE.7734E..2AP}, who showed a cutoff frequency above 220\,Hz. This is caused by the pure delays inside the system: detector integration time, processing time, and data transfer between the different units (LCUs).
The static gain of the piezo $G_{\rm piezo}^i=\sum_{k=1}^5 a_k^i$ is also an important property of the piezo-actuators because they are used both in the group- and phase-delay controllers.  Practically, to remove the static gain of the control loop, and because $\vec u$ is in unit of OPD radians instead of Volts, the $a_k^i$ values are normalized by the static gain of the $G_{\rm piezo}^i$.

\subsection{Observation equation}

The observable, $\vec \Phi_n$, depends on the state vectors of the piezo-actuators $\vec x_P$ and on the state vectors of the atmosphere and vibrations $\vec x_V$. Both contribute to the phase delay. The equation for the measurement process of the phase delay is a linear combination of the two:
\begin{equation}
\vec \Phi_{n}= \mathsf  M \cdot \mathsf  C_{P}   \vec x_{P,n} - \mathsf  C_{V}   \vec x_{V,n}  +\vec w_n + \mathbf{Ref} _{\vec \Phi}
\label{eq:erOPD}
,\end{equation}
where  $\vec w_n$ is a white measurement noise and $\mathsf   C_V$ and $\mathsf  C_P$ are three-dimensional measurement matrices with two-dimensional matrices on the diagonal for the vibrations and the piezo-actuator, respectively:
\begin{equation}
\mathsf C_V^{i,j}=\left( 
\begin{array}{cccccc}
1 &0 & 0 &\cdots&  0 & 0 \\
\end{array}
\right)
\end{equation}
\begin{equation}
\mathsf C_P^i=\left( 
\begin{array}{ccccc}
1 &0 & 0 & 0 & 0 \\
\end{array}
\right)
.\end{equation}

In the case of a resolved astronomical target, the phase-delay vector $\vec \Phi_{n}$ also includes a term that corresponds to the spatial signature of the target on the phase. It appears in the form of a non-zero closure phase ($\Theta^{\rm PD} \neq 0$) and is included in the phase delay by the term $\mathbf{Ref}_{\vec \Phi}$
, as defined in Eq.~(\ref{eq:RefPD}).

\subsection{Parameter identification}

The $a_{1\leq k \leq5}$ predictive terms for each of the piezo-actuators are determined offline during dedicated calibration laboratory measurements.  This is not the case for the identification of the $v_{1\leq k \leq30}$ AR values. They are calculated on a distinct Linux workstation. The workstation senses in real time the phase delay that passes by the reflective network. Every 5 seconds, it collects the last 40 seconds of data (36000 sample at 909\,Hz) and
 computes the pseudo-open-loop phase delay $\vec \Phi_{{\rm ATM},n}$ by removing the influence of the piezo command and closure phases. This phase delay corresponds to the phase delay produced by the atmosphere and the vibrations only:
\begin{equation}
\vec \Phi_{{\rm ATM},n}=  \vec \Phi_n - \mathsf  M \cdot \mathsf  C_{P}   \vec x_{P,n} - \mathbf{Ref} _{\vec \Phi} 
\label{eq:atm}
,\end{equation}
from which it calculates the differences between consecutive  values:
\begin{equation}
\vec \Delta \vec \Phi_{{\rm ATM},n}= \vec \Phi_{{\rm ATM},n}- \vec \Phi_{{\rm ATM},n-1}
.\end{equation}
The $\vec \Delta \vec \Phi_{{\rm ATM},n}$ differential phase values are then processed baseline after baseline to derive 29 $p_{1\leq i \leq29}$ AR parameters that best represent the data. This uses the Python toolbox "Time Series Analysis". The generated parameters ensure stationarity
\citep{JonesTSA} and thus provide stability to the closed-loop algorithm. Last, the $v_{1\leq i \leq30}$  values are determined through\begin{equation}
v_i=p_i-p_{i-1}
,\end{equation}
after which the matrices $\mathsf A_V$ and $\mathsf A_P$ are sent by the RMN to the real-time LCU to adjust the parameters of the phase-delay controller.

\subsection{Asymptotic Kalman filter}
\label{sec:kalFilter}

The vectors $\hat{ \vec x}_{V,n|n-1}$ correspond to our best estimate of the state of the vibrations at a moment $n$ from all the measurements available up to a time $n-1$. It can be estimated from $\hat{ \vec x}_{V,n-1}$ according to the equation of state in Eq.~(\ref{eq:state}):
\begin{equation}
\hat{ \vec x}_{V,n|n-1} = \mathsf  A_V \cdot \hat{ \vec x}_{V,n-1}\,.
\label{eq:stateEstim}
\end{equation}
However, the goal of the Kalman filter is to update this estimate from the error between the new observable and this new estimate. This error can be derived from the measurement process in Eq.~(\ref{eq:erOPD}) and writes
\begin{equation}
\vec \varepsilon_{\vec \Phi,n}= \mathbf{Ref}_{\vec \Phi}  - \vec \Phi_{n} + \mathsf  M \cdot \mathsf  C_{P}   \vec x_{P,n} - \mathsf  C_{V}    \cdot \hat{ \vec x}_{V,n|n-1}    \,, 
\label{eq:varpd}
\end{equation}
which becomes, after unwrapping between $-\pi$ and $\pi$ and baseline weighting,
\begin{equation}
\vec {\varepsilon '}_{\vec \Phi,n} = \mathbb I_{\rm PD}^6.  ( \left[ \vec \varepsilon_{\vec \Phi,n} \right] \%  2 \pi  ) \,,
\label{eq:eI6}
\end{equation}
where $\%  2 \pi$ corresponds to modulo $2\pi$. The 
$\mathbb I_{\rm PD}^6$ matrix is here for two purposes. The first is to derive the errors on low S/N baselines through the baselines with higher S/N. The second purpose is
to ensure that the low S/N data are either weighted down or  discarded through the weighted pseudo-inversion in Eq.~(\ref{eq:sinvPD2}).
The state estimator $\hat{ \vec x}_{V,n}$ is finally updated through an integrator: 
\begin{equation}
\hat{ \vec x}_{V,n} =  \mathsf  A_V \hat{\vec x}_{V,n-1} + \mathsf G_{\rm PD} \cdot  \vec {\varepsilon '}_{\vec \Phi,n} \label{eq:G} \,.
\end{equation}

The gain $\mathsf  G_{\rm PD}$ is not a scalar but a three-dimensional matrix. The correct estimate of  $\mathsf  G_{\rm PD}$  is the basis of Kalman filtering. In GRAVITY, it is not identified at each DIT, but every 5 seconds on the sensing Kalman workstation. Therefore, it is obtained from the asymptotic Kalman equations. It is calculated from the two covariance matrices of the measurement noise $\Sigma_w$ and the steady-state error  $\Sigma_\infty$ \citep{2010JOSAA..27A.223P,2012A&A...541A..81M}:
\begin{equation}
\mathsf  G_{\rm PD}= \Sigma_\infty \mathsf  C_V^\dag ( \mathsf  C_V \Sigma_\infty \mathsf  C_V^\dag + \Sigma_w)^{-1}
.\end{equation}
The steady-state covariance matrix can be obtained from the algebraic Riccati equation, the vibrations input noise, and the vibration and atmospheric noise $\Sigma_v$:
\begin{equation}
 \Sigma_\infty = \mathsf  A_V  \Sigma_\infty \mathsf  A_V^\dag - \mathsf  A_V \Sigma_\infty  \mathsf  C_V^\dag ( \mathsf  C_V \Sigma_\infty \mathsf  C_V^\dag + \Sigma_w)^{-1} \mathsf  C_V \Sigma_\infty  \mathsf  A_V^\dag +\Sigma_v
.\end{equation}
The noise characteristic $(\Sigma_v,\Sigma_w)$ makes the Kalman filter optimally adapted to the average noise on the system, but not to instantaneous noise variations. Adaptability is the role of the $\mathbb I_{\rm PD}^6$ matrix in Eq.~(\ref{eq:eI6}).

\subsection{Determination of the control signal}
\label{sec:kalCommand}

To use the Kalman filter as a controller, an optimal command is needed. The purpose of this section is to determine the control signal $\vec u_{ \textrm{PD}, n}$ from the atmospheric state vectors: $ \vec u_{ \textrm{PD}, n} =  \mathsf  K \hat {\vec x}_{V}$.
The task of this signal is twofold. First, it must cancel the atmospheric perturbations on the measured phase delay ($\vec \Phi_{{\rm ATM},n}$). Second, it must ensure that
 the phase delay converges to the group-delay control signal $\vec u_{\mathrm{GD},n} + \vec u_{\mathrm{modulation}}  + \vec u_{\mathrm{search},n}$. 
 
 The first task uses the predictive power of the equation of state:
\begin{equation}
  \vec u_{  \textrm{PD}, n} = \mathsf  M^\dag \mathsf  C_V \mathsf  A_V^{n_{\rm DIT}}   \hat {\vec x}_{V}
  \label{eq:ucompd}
,\end{equation}
where $\mathsf  A_V^{n_{\rm DIT}}$ is the power of matrix $\mathsf A_V$ over $n_{\rm DIT}$ samples. The integer ${n_{\rm DIT}}$ is there to account for the delay between control signal and its effect on the phase delay. Ideally, ${n_{\rm DIT}} =1,$ but because of the pure delay in the open-loop transfer function, it must be higher. This pure delay depends on the frequency of the fringe tracker. We currently use  $n_{\rm DIT} =3$ for the 909\,Hz frequency and $n_{\rm DIT} = 2$ for the lowest frequencies.

The second task  is achieved through the integrator in Eq.~(\ref{eq:G}).  This integrator updates the state vector  of the atmospheric perturbations to minimize the error between estimation and observation. It causes the quantity $\vec \varepsilon_{\vec \Phi,n}$ to converge to zero: $\vec \varepsilon_{\vec \Phi,n \rightarrow \infty} = 0$.   From Eq.~(\ref{eq:varpd}), we can derive
\begin{equation}
 \vec \Phi_{n \rightarrow \infty} = \mathbf{Ref}_{\vec \Phi}  + \mathsf  M \cdot \mathsf  C_{P}   \vec x_{P,n \rightarrow \infty} - \mathsf  C_{V}    \cdot \hat{ \vec x}_{V,n \rightarrow \infty}    \, .\end{equation}
 Under a steady-state assumption, $ \mathsf  C_{P}   \vec x_{P,n \rightarrow \infty} = \vec u_{n \rightarrow \infty}$ because the piezo-gains are normalized. From Eq.~(\ref{eq:ucompd}) we also have, under the same assumption,
  $\mathsf  C_{V}    \cdot \hat{ \vec x}_{V,n \rightarrow \infty} =  \mathsf  M \cdot\vec u_{\mathrm{PD},n \rightarrow \infty}$ because the first row of $\mathsf  A_V$ is normalized. Hence, we have what we desire, i.e.,
 \begin{eqnarray}
          \vec \Phi_{n \rightarrow \infty}                 &=& \mathbf{Ref}_{\vec \Phi}  + \mathsf  M \cdot ( \vec u_\infty - \vec u_{\mathrm{PD},n \rightarrow \infty} ) \\
                         &=& \mathbf{Ref}_{\vec \Phi}  + \mathsf  M \cdot ( \vec u_{\mathrm{GD},n \rightarrow \infty} + \vec u_{\mathrm{modulation}}  + \vec u_{\mathrm{search},n \rightarrow \infty}) \,.\end{eqnarray}


   \begin{figure*}
   \centering
   \includegraphics[width=0.8\textwidth]{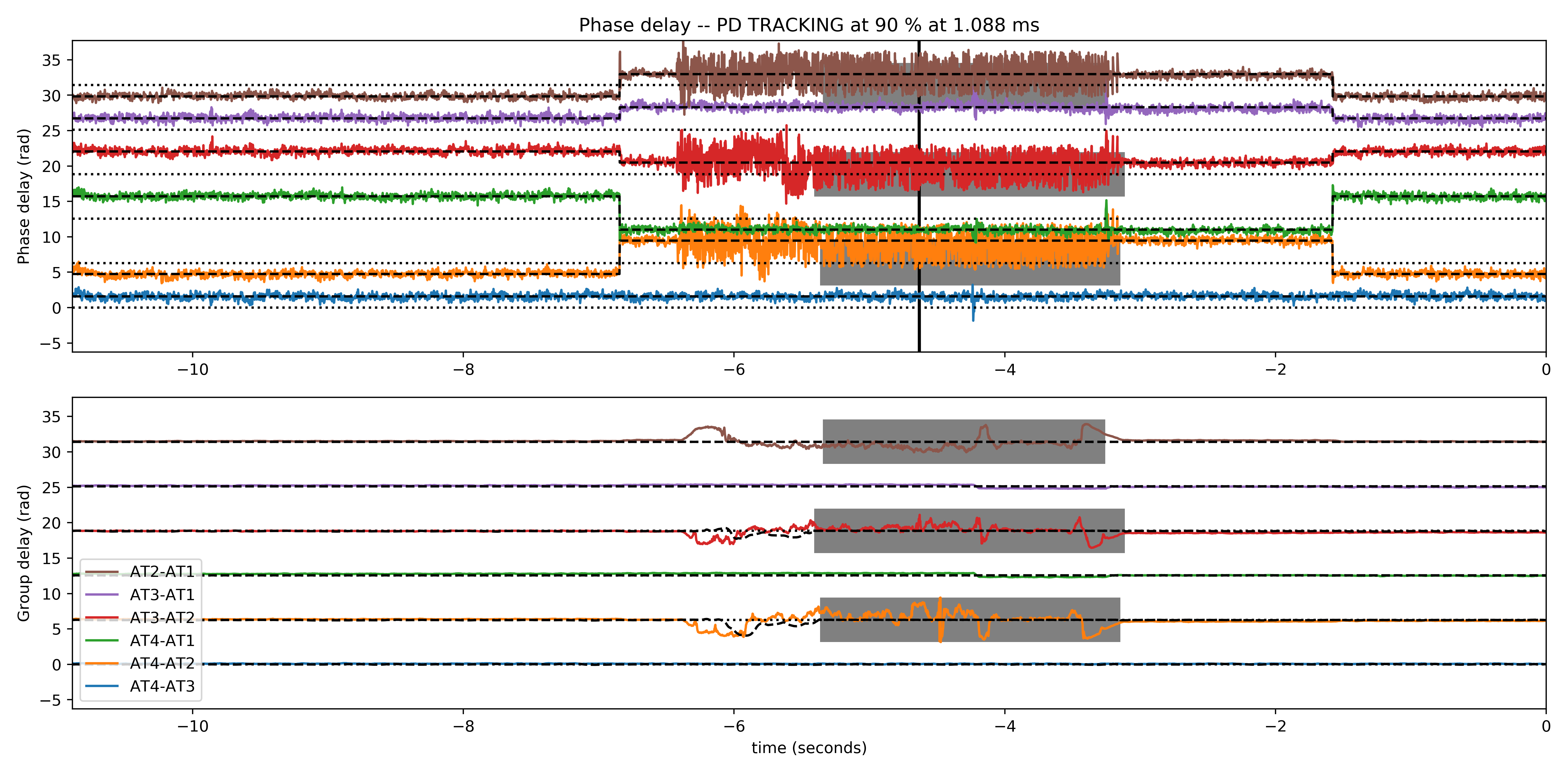}
   \includegraphics[width=0.8\textwidth]{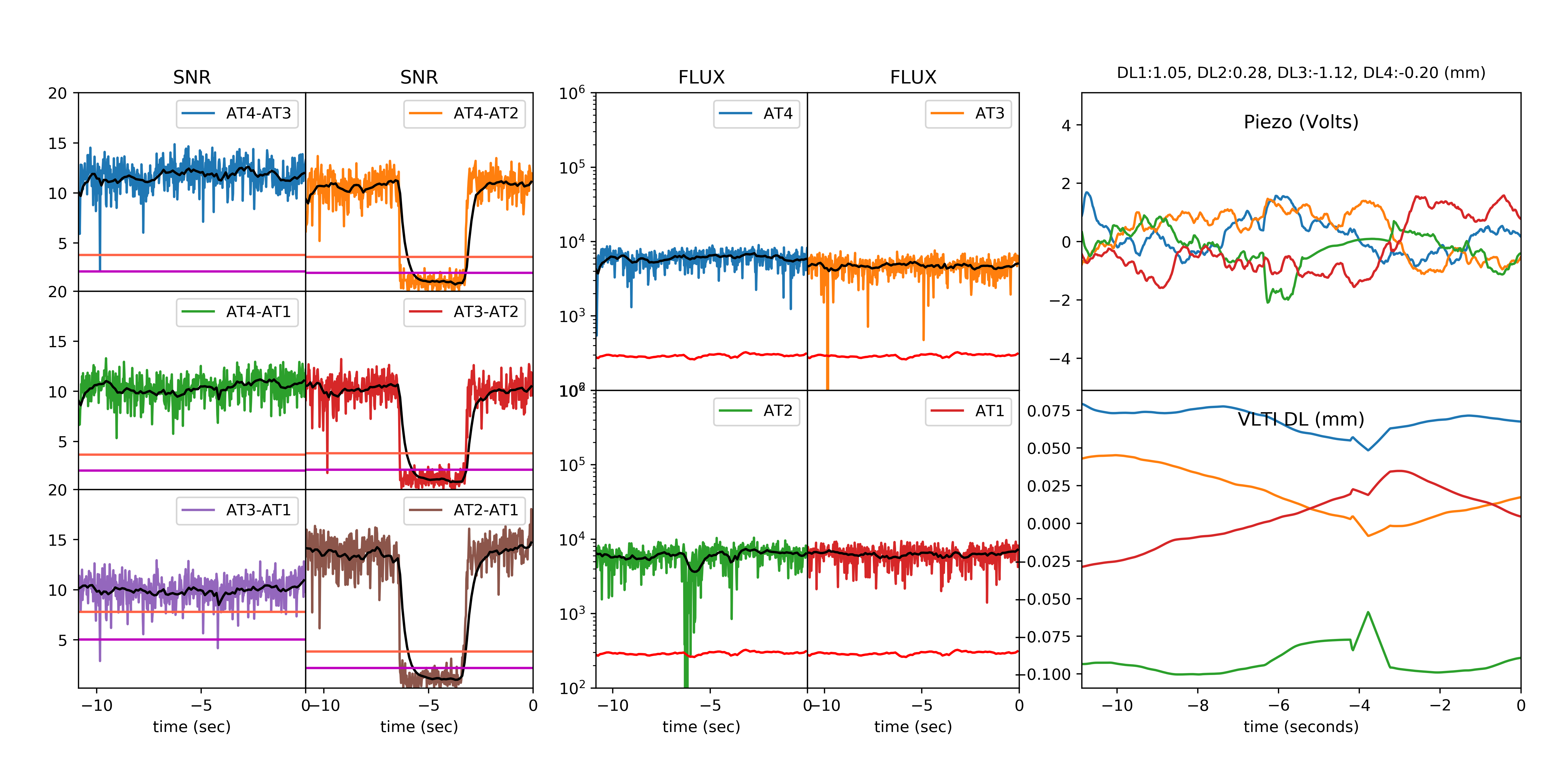}
   \caption{Observation of star GJ\,65\,A on 21 November 2018.  This example shows the case of a glitch on the AT2 adaptive optics system that resulted in the loss followed by the recovery of the fringe by the fringe tracker. \textbf{Upper two panels:} Fringe tracker phase  $\Phi_{i,j}$ and group delay $\Psi_{i,j}$ estimators. 
   The color lines correspond to each of the six baselines. 
   The $\pi/2$ phase jumps at -7 and-1.5 seconds are normal and correspond to the modulation synchronized with the 5s DIT science camera (the high spectral resolution detector). The gray areas correspond to detection of low S/N fringes by the controller.   
   \textbf{Lower left panels:} S/N for each of the baselines calculated as the inverse of the square root of the phase variance ($1/\sqrt{Var(\Phi_{i,j})}$). The horizontal red lines correspond to the group-delay threshold ($s/n_{\rm threshold}^{\rm GD}$). If the black line dips below the red line, the FT stops tracking on that baseline. The horizontal purple lines correspond to the phase-delay threshold ($s/n_{\rm threshold}^{\rm PD}$).    \textbf{Lower center panel:}  Flux $F_i$. Similarly, the red line is a threshold made by a moving average of all flux, which is used to detect the loss of a telescope. \textbf{Lower right panels:} Commands to the piezo-actuators and VLTI delay line actuators. The piezo-actuators take care of the fast control signal, while the VLTI delay lines are used to offload and to search for fringes over large distances.
   }
              \label{fig:loss}%
    \end{figure*}

\subsection{Why not a simpler state controller?}

The block diagram presented in Fig.~\ref{fig:control} shows the complexity of the fringe tracker. A simpler version  was used during the first commissioning in 2016. The phase and group-delay controllers were both different. First, the group-delay control signal was not used to directly command the piezo-actuators, but instead, it was used as a setpoint for the phase-delay controller. Second, the phase-delay controller was a proportional-integral controller. It was efficient and robust and a good commissioning tool. However, at low S/N, its performance was not adequate. The reasons were that
\begin{itemize}
\item the phase-delay estimator is noisy. The proportional controller (or worse, a derivative) directly injected the noise back into the control signal. This was especially problematic during flux dropouts when the S/N can be close to zero; and that
\item the phase delay is modulo $2\pi$. It needs to be unwrapped with respect to a prediction. Using the setpoint as the prediction resulted in many fringe jumps during poor atmospheric conditions.
\end{itemize}
We therefore decided to have the group delay directly command the piezo, skipping the phase-delay controller. In parallel to the group-delay controller, we included a Kalman filter on the phase-delay feedback and used the predictive model to unwrap the phase (the vector $\vec {\varepsilon '}_{\vec \Phi,n}$). This gave the block diagram in Fig.~\ref{fig:PDblocks}.

       
\section{On-sky observations}
\label{sec:results}

\subsection{Operation}

The simplicity of operating the fringe tracker lies in the simplicity of the state machine. With only three states, the operator interactions are limited to the transition between IDLE, SEARCH, and back to IDLE. When the fringes are detected, the system automatically switches to TRACKING and the instrument then starts recording data. 
However, there are two free parameters that could ask for the intervention of the operator: the $s/n_{\rm threshold}^{\rm GD}$ and $s/n_{\rm threshold}^{\rm PD}$ thresholds. An operational error could be, for example, to set 
a $s/n_{\rm threshold}^{\rm GD}$ too low and risk having the fringe-tracker tracking on the second lobe of a fringe packet.

When the thresholds are correctly set, the system is made to be fully autonomous, and able to deal with any glitch of the VLTI. For example, Fig.~\ref{fig:loss} shows the fringe tracker losing telescope AT2 and how it recovered. The figures are the same plots as those available to the support astronomers during real-time GRAVITY operation. The data were captured and plotted at $t=0$\,s.
At $t=-6.39$\,s, the AT2 looses its pointing target, and the injected flux in the fiber drops. Immediately, the S/N level drops below the purple line, and the FT discontinues tracking on all the three baselines that include AT2. At this point, the system is still in TRACKING state. At $t=-5.39$\,s, after a time delay of 1 second, the system switches to SEARCHING state, and the VLTI delay lines start following the sawtooth function. 
At $t=-3.26$\,s, the system recovers the fringes on the AT2-AT1 baseline and starts centering them. At this point, the rank of the $\mathbb I_{\rm GD}^4$ matrix is back to 3 and the fringe tracker  switches back to TRACKING state. At $t=-3.12$\,s, the fringes are detected on all six baselines, and the system again tracks nominally.

In summary, the complexity of dealing with multiple baselines is hidden behind the $\mathbb I_{\rm GD}^6$ and $\mathbb I_{\rm PD}^6$ matrices presented in Sec.~\ref{sec:I}. From the user's point of view, the fringe tracker transitions from SEARCHING to  TRACKING state, but the engine behind the scene does not change the way it operates.

\subsection{Sensitivity}

   \begin{figure}
   \centering
   \includegraphics[width=0.48\textwidth]{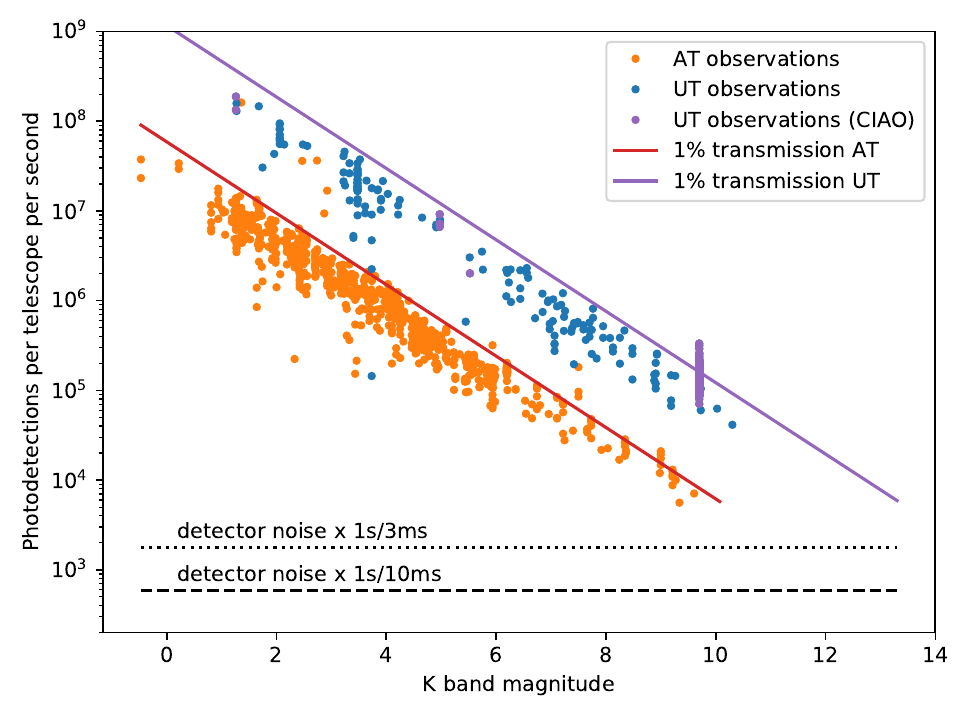}
   \caption{Transmission plot, i.e., photons detected on the GRAVITY FT receiver, per telescope and per second, as a function of the K-band magnitude (tracking ratio $>80\%$). The magnitudes of the targets are obtained from SIMBAD. Most of the observations are made on-axis, meaning that 50\% of the flux is lost because of the beam splitter.   
  However, the many K=9.7\,mag CIAO observations are taken off-axis, hence the higher flux.  Targets observed with ATs can be as faint as 10 magnitudes. Based on this plot, it is clear that because the flux observed with the UTs is more than 10 times higher, observations will be possible up to K=13 mag. }
              \label{fig:Photons}%
    \end{figure}
    
   \begin{figure}
   \centering
   \includegraphics[width=0.48\textwidth]{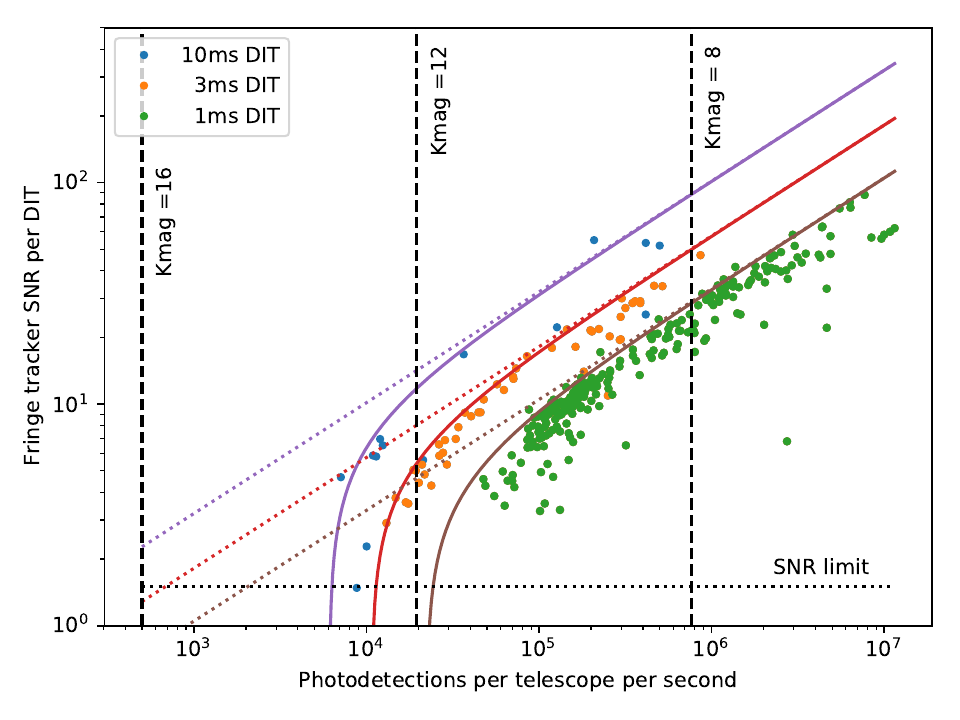}
   \caption{Signal-to-noise ratio as a function of detected photon per telescope per DIT. The S/N is the computed by the real-time computer, defined by $1/\sqrt{Var(\Phi_{i,j})}$ , as stated in Eq.~(\ref{eq:varPhi}). The three colors correspond to the three different frame rates of the fringe tracker. The solid lines correspond to theoretical values assuming 100\,\% fringe contrast. The vertical lines are theoretical flux assuming observations on UTs with 1\% throughput. The horizontal line corresponds to an S/N of 1.5.
   }
              \label{fig:SNR}%
    \end{figure}

The sensitivity is mostly a question of having enough photons on the detector to generate a feedback signal for the fringe tracker. In Fig.~\ref{fig:Photons} we plot the flux versus magnitude of calibrators observed with GRAVITY during a period covering June 2017 to November 2018.
The selection of the files with a tracking ratio above 80\%  leads to a total of 1117 exposures on 473 distinct calibrators.  The orange dots correspond to the 814 AT observations. The others correspond to the MACAO (visible AO) and CIAO (infrared AO) UTs observations. The solid lines correspond to a total transmission of 1\% from telescope to the fringe-tracker detector. The dashed lines correspond to the theoretical detector noise, scaled by $\sqrt{N_p}$, with $N_p$ the total number of pixels divided by the number of telescopes. 
    
On the ATs,  the faintest target  observed so far was TYC\ 5058-927-1, a star with $K = 9.4$\,mag. It was observed in the night of 5 April 2017, when the atmospheric conditions were excellent: seeing down to 0.4'' and coherence time up to 12\,ms. The observations were made in single-field mode, where only half the flux is sent to the fringe tracker. The fringe tracker efficiently tracked the fringes throughout the entire exposure time, at a frequency rate of 97\,Hz, with OPD residuals between 250 and 350\,nm.

Technically, if we were to extrapolate to the UTs the sensitivity observed on the ATs, we should be able to track stars of K magnitudes up to 12.5. However, the faintest star observed with the fringe tracker on the UTs so far is TYC\ 7504-160-1, a star with $K = 11.1$\,mag. It was observed in the night of 2 July 2017 during good atmospheric condition: seeing between 0.4’’ and 0.6'', coherence time between 4 and 7\,ms. The frequency rate was 303\,Hz, with OPD residuals between 350 and 400\,nm.
The relatively low UT sensitivity is still not understood. One explanation could be that the 97\,Hz integration time  cannot be used: the fringe contrast is attenuated by the vibrations of the UT structure. Another explanation could come from moderate AO performances. With a low Strehl ratio, difficulties in injecting the light in the fibers decrease the number of available photons, but also create flux drops that decrease the visibility and complicate the fringe tracking. One last possibility is a selection effect caused by the lower availability of the UTs.

\subsection{Signal-to-noise ratio}

The limit for fringe tracking could in theory be an S/N of 1 per coherence time. Below that number, the phase varies faster than our ability of measuring it: our knowledge  of the phase decrease with time, hindering the convergence of the fringe tracker. However, practically, we observed that an S/N of 1.5 per DIT is required to keep the fringes in the coherence envelope. 
The signal is proportional to the number of coherent photons received.
The noise is created by quantum noise on one hand and the background noise on the other hand.  At 909, 303, and 97\,Hz, we measured during sky observations a standard deviation of $\sigma_{\rm sky}=4$, 5, and 8\,ADU per pixel, respectively. 
These values come from the quadratic sum of the noises caused by the scattered metrology light (7 ADU at 97\,Hz), the sky and environmental background (6 ADU at 97\,Hz), and the read-out noise (4\,ADU). 

The detector variance observed during sky observations is used to compute the real-time S/N shown in the lower left panel of Fig.~\ref{fig:loss}. 
In Fig.~\ref{fig:SNR}, for a dataset covering the seven months between April and November, we have plotted this S/N as a function of the measured flux. 
The three solid lines correspond to the theoretical S/N calculated from $\sigma_{\rm sky}$ and photon noise. 
These values lie below these theoretical lines because of a loss of visibility contrast. This can be caused either by non-equilibrium of the flux between the different telescopes, or by OPD variations within a DIT of the fringe tracker detector.

The vertical lines correspond to the flux of a star of K magnitude of 8, 12, and 16 observed with the UTs under the assumption of 1\% throughput. 
The drop at low flux of the theoretical S/N curves correspond to the effect of the observed sky noise. This noise is mostly  detector noise at high frequency, and a combination of background and metrology noise at low frequency. 
The dotted lines are theoretical computations of the noise assuming only photon noise. Under this assumption, 1\% throughput, and 100\% visibilities, the UT sensitivity could technically reach magnitudes up to $K=16\,$mag.

\subsection{OPD residuals and S/N}

   \begin{figure}[t]
   \centering
   \includegraphics[width=0.48\textwidth]{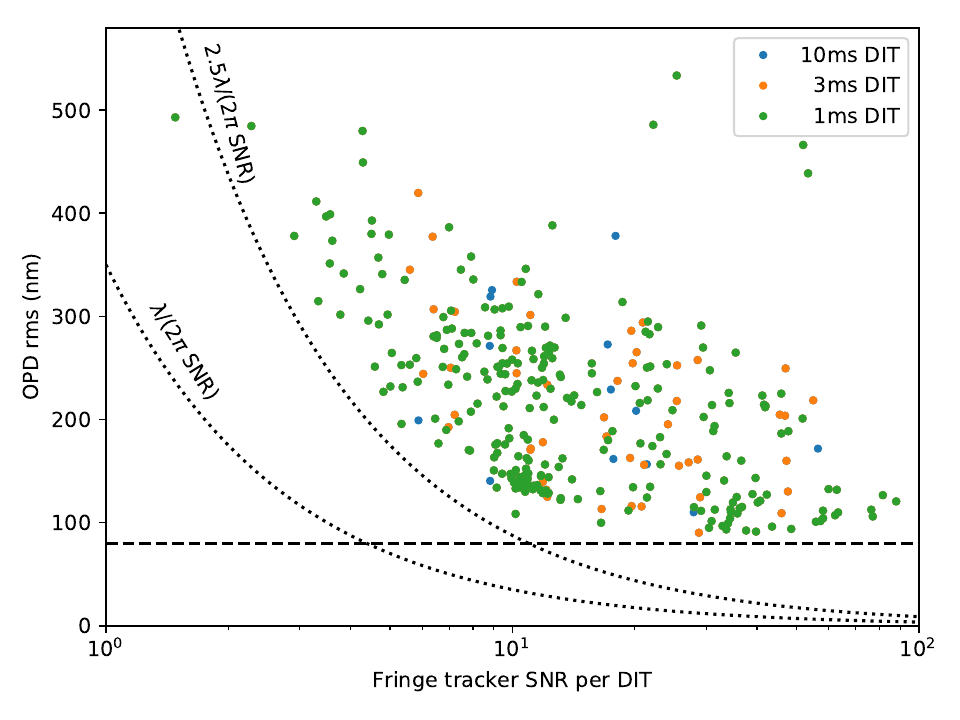}
   \caption{OPD residuals as a function of the fringe tracker S/N. Above an S/N of 10, the best fringe-tracking residuals are limited to a constant value around 80\,nm, caused by the intrinsic bandpass of the fringe tracker hardware (dashed line). At lower S/N, the fringe tracker performance decreases because it is difficult to predict and track the  evolution of the perturbations. The phase error caused by a readout noise on the phase goes as  $\lambda/(2\pi\, {\rm S/N})$ (lower dotted line). The actual performances of the fringe tracker shows that the degradation is more likely a factor 2 or 3 above that limit.}
              \label{fig:OPDrms_snr}%
    \end{figure}

The fringe tracker performance degrades when the limiting sensitivity is approached. This is partly because the phase-delay control-loop gain decreases at low S/N because of the weighted  inversion of matrix $ \mathsf S^{\dag_{\rm PD}} $ in Eq.~(\ref{eq:sinvPD2}). This is also partly caused by the difficulty of estimating the correct fringe-tracking state from a noisy measurement. This decrease in performance is shown in Fig.~\ref{fig:OPDrms_snr} for the same dataset as presented in Fig.~\ref{fig:SNR}. Concretely, below a S/N of 4, the residuals are above 300\,nm. These residual include the variance caused by the measurement noise, which is equal to $\lambda/(2\pi\, {\rm S/N})$: $\approx 90\,$nm for a S/N of 4. The fringe tracker residual does not correct the phase to that level, however: the residual remains a factor 2 or 3 above that value. At an even lower S/N (<3), the Kalman filter has problems to identify the state of the atmosphere and sometimes cannot update them fast enough. Fringe tracking is then only possible when the atmosphere varies slowly, when the atmospheric conditions are best.
    
\subsection{OPD residuals and $\tau_0$}

   \begin{figure}
   \centering
   \includegraphics[width=0.48\textwidth]{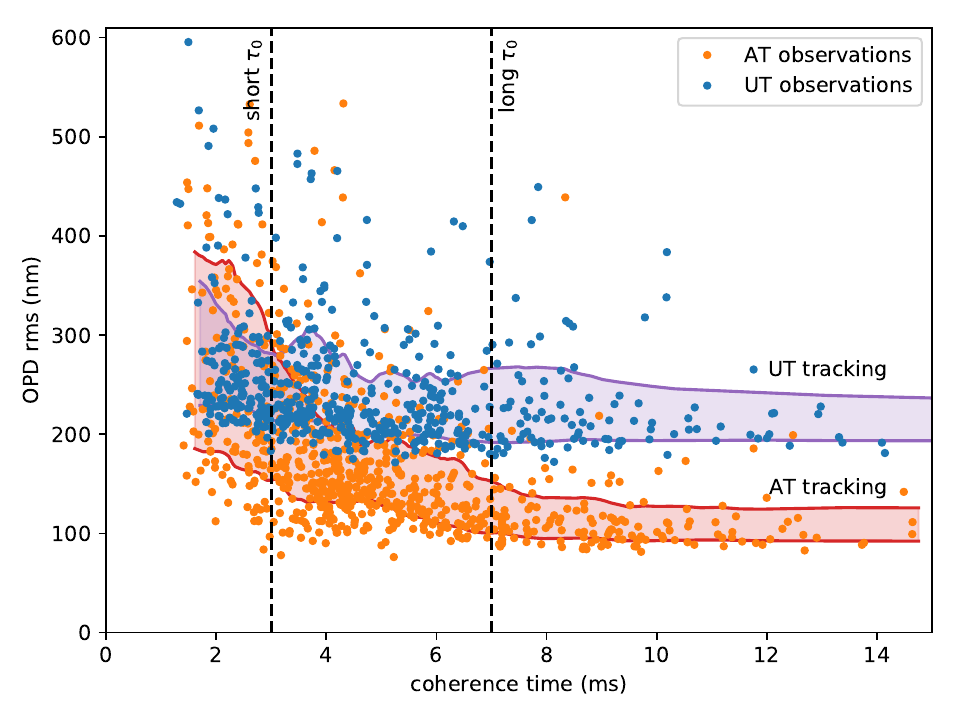}
   \caption{OPD residuals as a function of coherence time (at 500\,nm) for all calibrators and GC data taken between July 2017 and November 2018 in the 909\,Hz mode. The dashed vertical lines correspond to the first quintile (3\,ms) and last quintile (7\,ms) of the coherence time distribution as observed by the ESO astronomical site monitor (ASM) at visible wavelength. The thick lines correspond to the first and last quintile of the OPD residual distribution (purple for UTs and red for ATs). The median fringe-tracking residuals are 150\,nm on the ATs and 250\,nm on the UTs.}
              \label{fig:OPDrms}%
    \end{figure}

   \begin{figure*}
   \centering
   \includegraphics[width=\textwidth]{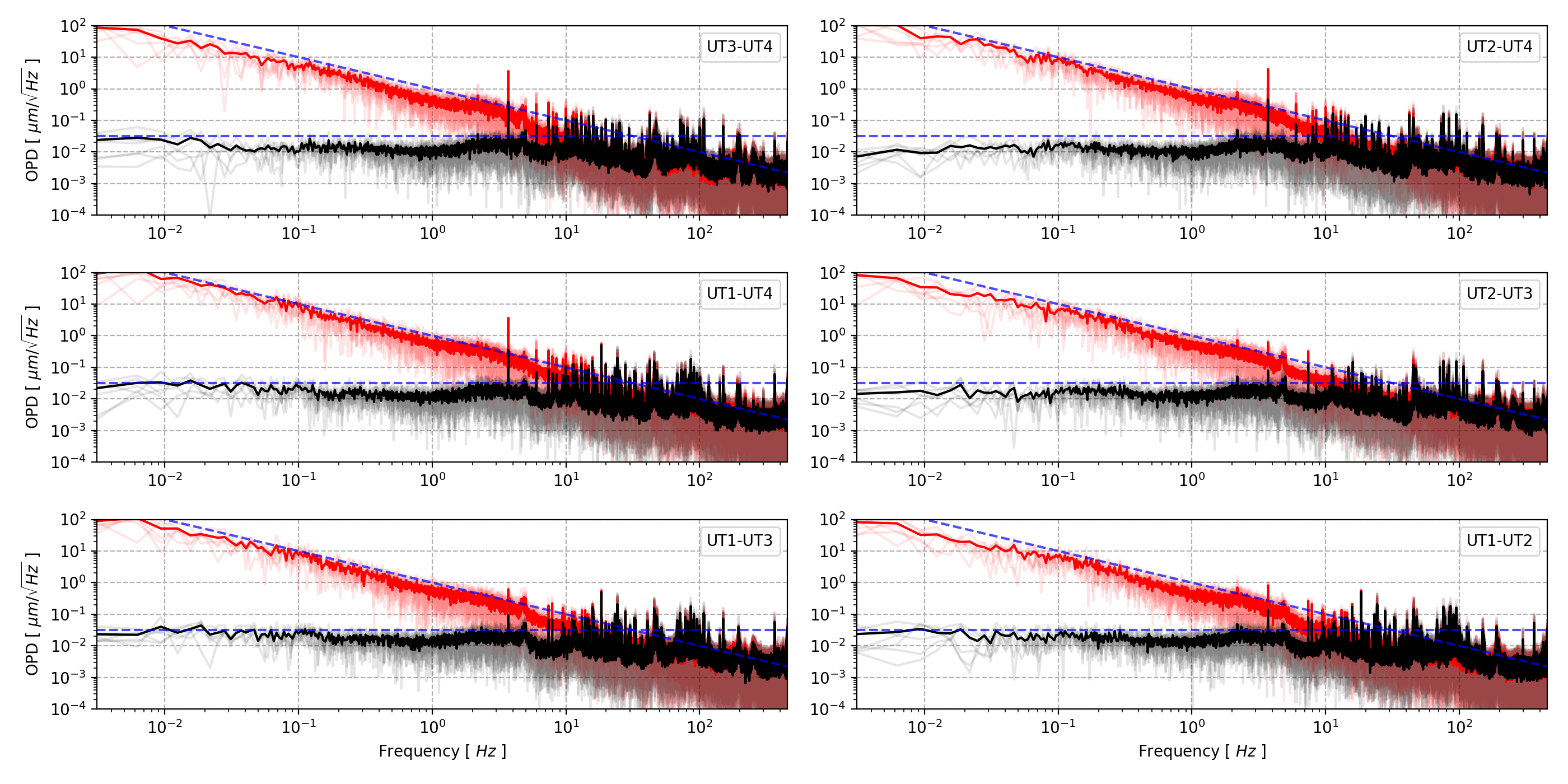}
   \includegraphics[width=\textwidth]{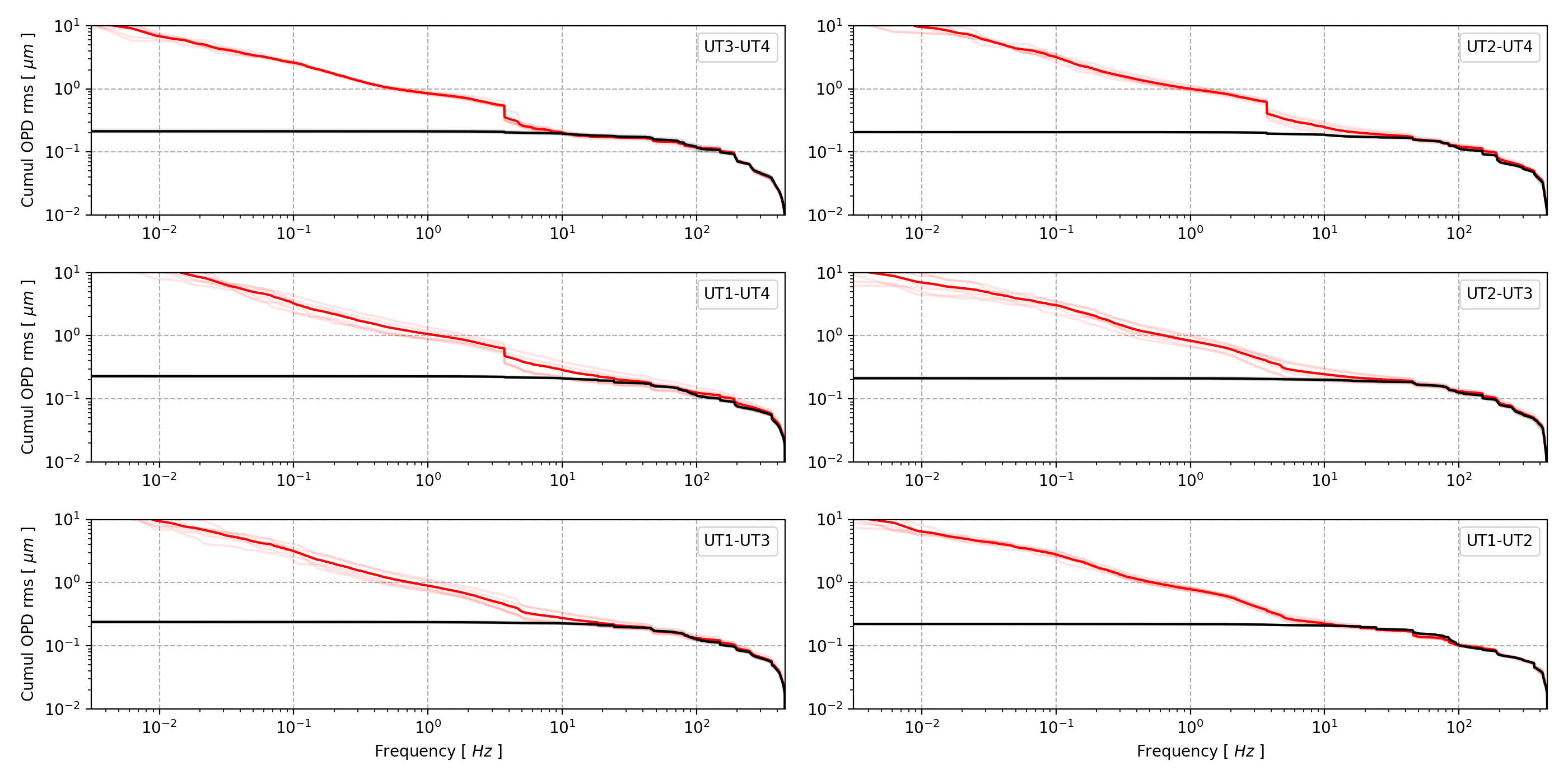}
   \caption{Power spectrum density and cumulative sum of the power spectrum density of phase residuals ($2.2\,\mu$m$/2\pi \times \vec \Phi_{n}$) toward target IRS\,16C. The red curves are the pseudo-open-loop values, i.e., in the hypothesis of no fringe tracker ($2.2\,\mu$m$/2\pi \times \vec \Phi_{{\rm ATM},n}$).
    The blue curves  correspond to power spectral densities of $
1\,f^{-2} \, \mu{\rm m}^2/{\rm Hz}$ and $
 0.001 \, \mu{\rm m}^2/{\rm Hz}$. The six lower plots are the reverse cumulative sum of the power spectrum. At zero frequency, the sum reaches 220\,nm on average over the six baselines.
   }
              \label{fig:PSDgc}%
    \end{figure*}
    
 At high S/N ($\leq 10$), the accuracy of the fringe tracking depends on several environmental parameters: vibrations, wind speed, seeing conditions and coherence time ($\tau_0$). Between 
wind seeing and coherence time, \citet{2018SPIE10701E..07L} showed that the strongest correlation is observed with $\tau_0$.
Using the same dataset as in Fig.~\ref{fig:Photons} and using only the calibrators observed at 909\,Hz,  we plot in Fig.~\ref{fig:OPDrms} 
the OPD residuals as a function of the coherence time as observed by the ESO site monitor at 500\,nm. On average, the ATs perform better, with a median residual of 150\,nm. Under the best conditions, the OPD residuals can be as low as 75\,nm. The UTs residuals are higher, with a median value of 250\,nm, and a minimum at 180\,nm.

Depending on the seeing conditions, the fringe tracker shows different limitations. In the worst atmospheric conditions, which make up 20\%,\ ($\tau_0<3$\,ms), the UT and AT performances are limited by the  coherence time of the atmosphere. Under these conditions, 20\% of all observations have OPD residuals above 380\,nm. The consequence is then that 
 the jitter of the phase significantly affects the visibility in the science channel with long integration times. We can show that this contrast loss can be directly estimated from the variance of the phase according to the relation 
\begin{equation}
V_{\rm residuals}= \langle \exp \, ( i \Phi ) \rangle= \exp \, (-\sigma^2_\Phi/2) \,.
\end{equation}
With 380\,nm rms jitters ($\sigma_\Phi \approx 1.1\,$rad), the contrast of the fringes of the science detector will only be at $V_{\rm residuals}\approx55\%$ of its maximum.

During the best 20\% of atmospheric conditions ($\tau_0>7$\,ms), however, the fringe tracker can reach very low residuals. On the ATs, 80\% of all observations are below 140\,nm ($V_{\rm residuals}>93\%$). However, on the UTs, the environmental vibrations  cause the residuals to remain between 220 and 320\,nm. The explanation is the higher level of vibrations observed on the UTs.

   \begin{table}
      \caption[]{GRAVITY dataset plotted in Fig.~\ref{fig:PSDgc} and \ref{fig:PSDgj65} }
\label{table:GRdataset}      
\centering                          
\begin{tabular}{c c c c c}        
\hline\hline                 
            Object   &  MJD & FT freq. & Seeing & $\tau_0$ \\
             Configuration& & K\,mag & & (ms) \\
\hline                        
   IRS\,16C & 58209& 909\,Hz  & 0.4''  & 0.007\\
    UT1-UT2-UT3-UT4 &  & 9.7\\      
\hline                        
    GJ\,65\,A (\&B) &  58367& 909\,Hz  & 0.6'' & 0.006 \\
    A0-G1-J2-K0 & & 5.7 (5.9) \\
\hline                                   
\end{tabular}
\end{table}

\subsection{Power spectral density}

   \begin{figure*}
   \centering
   \includegraphics[width=\textwidth]{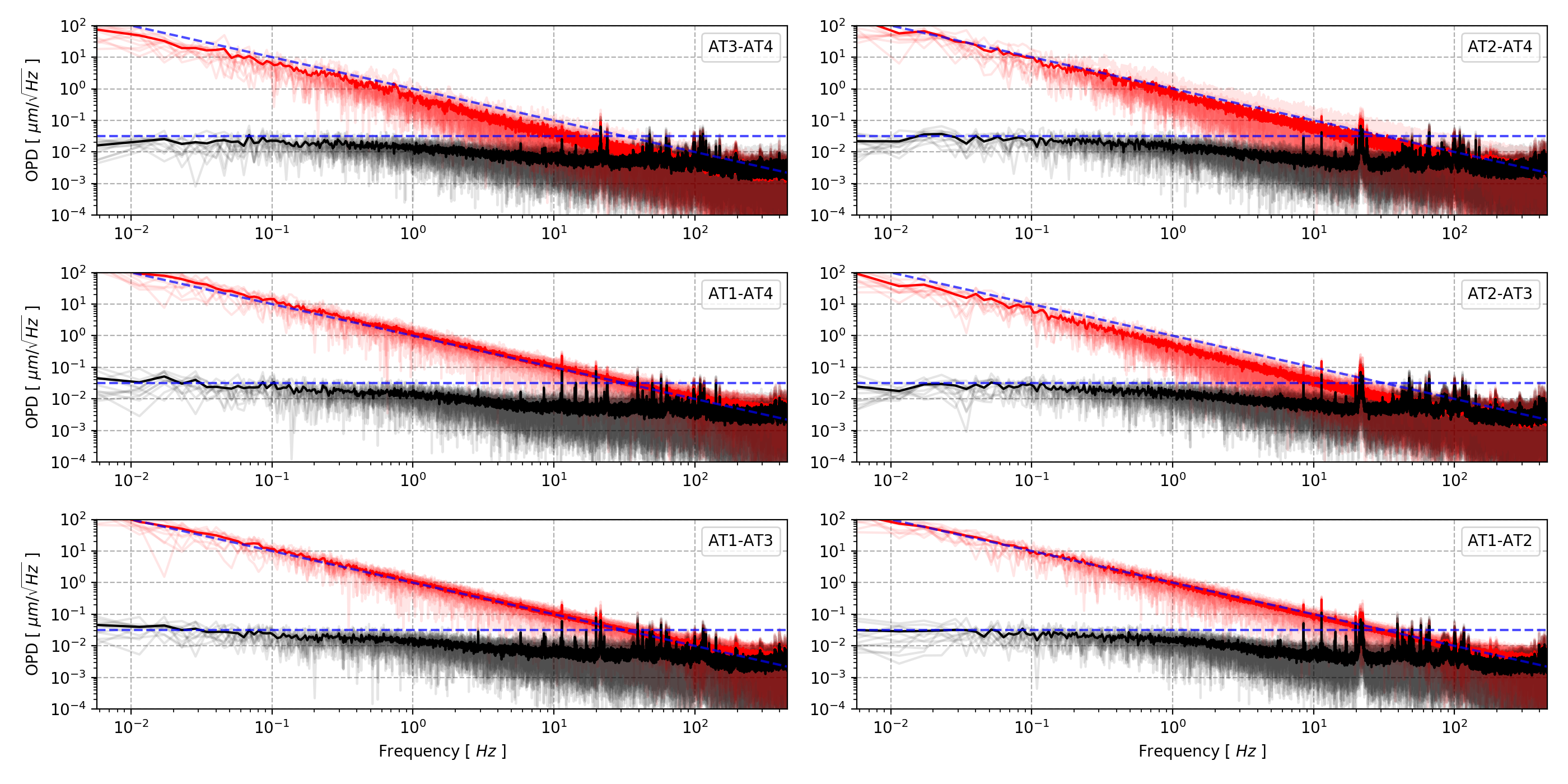}
   \includegraphics[width=\textwidth]{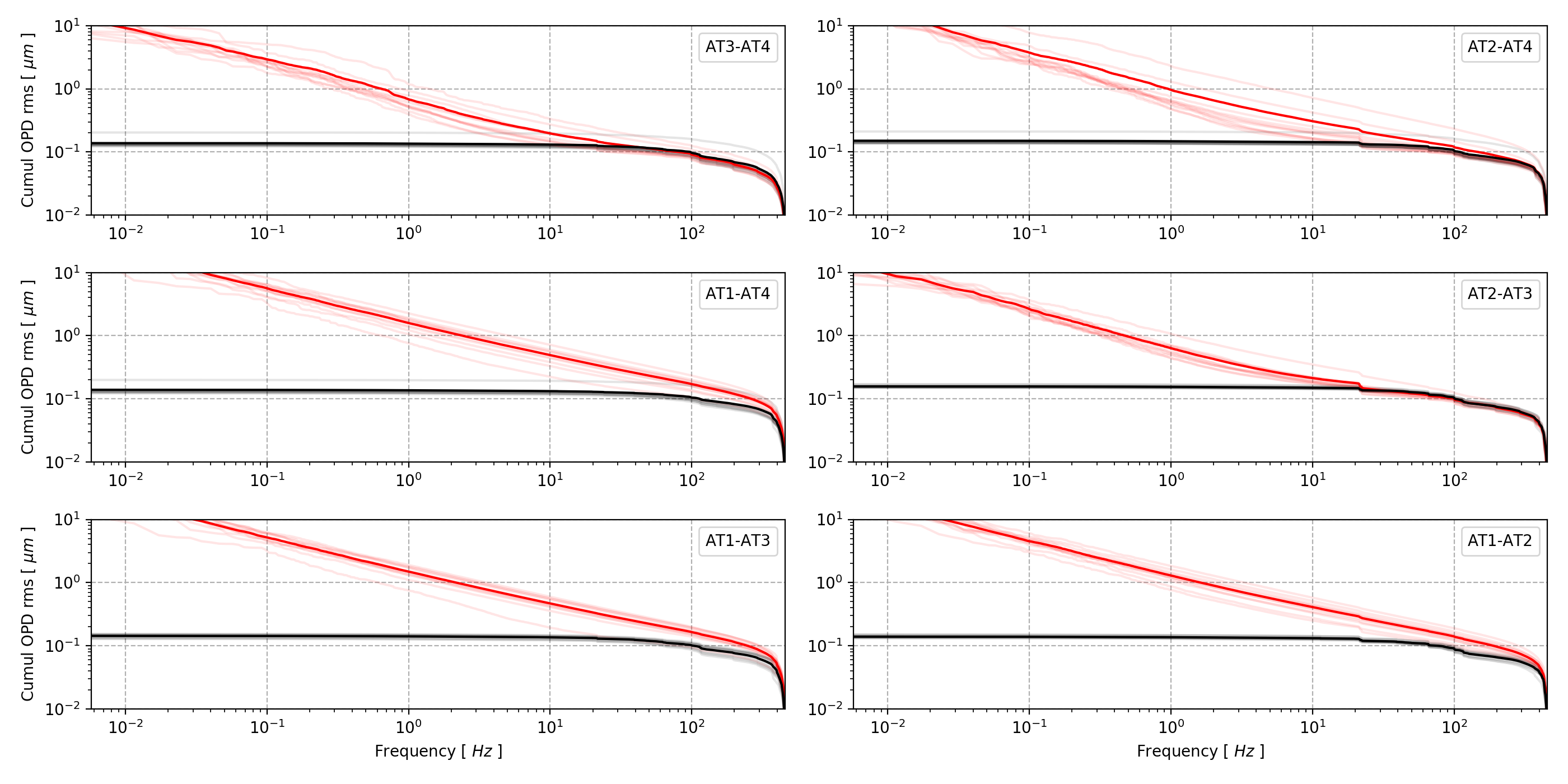}
   \caption{Same as in Fig.~\ref{fig:PSDgc}, except for the binary star GJ\,65, which was observed with the ATs. The fringe-tracking residuals are 140\,nm, lower than what is observed with the UTs. The main difference is the lower level of vibrations in the 10-100\,Hz range.}
              \label{fig:PSDgj65}%
    \end{figure*}

In Figs.~\ref{fig:PSDgc} and~\ref{fig:PSDgj65} we show the power spectral density of the OPD for two astronomical objects: IRS\,16C observed with UTs, and GJ\,65 observed with ATs.
The observation conditions are listed in Table~\ref{table:GRdataset}. The IRS\,16C galactic center target was observed with the CIAO off-axis AOs. The conditions were good to excellent, and the residual OPD was around 220\,nm over the entire duration of the observations. The GJ\,65 binary was also observed in good seeing conditions, without AO, and OPD residuals of 140\,nm.

To compute the power spectral density, the phase delay $\vec \Phi_n$ is unwrapped and the $\pi/4$ modulation function removed. The pseudo-open-loop phase delay $\vec \Phi_{{\rm ATM},n}$ is then computed from Eq.~(\ref{eq:atm}) using the piezo transfer function estimated during calibration. In Figs.~\ref{fig:PSDgc} and~\ref{fig:PSDgj65}, the six upper plots correspond to the square root of the  power spectral density. The black curves correspond to the measured OPD estimated from the phase delay $\vec \Phi,$ and the red curves show the power spectrum of the reconstructed atmosphere  $\vec \Phi_{{\rm ATM},n}$.
The atmospheric power spectrum (which also includes the vibrations) shows, as expected, the prevalence of the low frequencies in the atmospheric perturbations. Below 10\,Hz, the power spectral density fits the relation\begin{equation}
{\rm PSD_{ \vec  \Phi_{{\rm ATM},n}}}(f<10\,Hz)=1f^{-2} \, \mu{\rm m}^2/{\rm Hz}\end{equation}
very well, which is not far, but different, from the power of -5/3 of a Kolmogorov type atmosphere. After fringe-tracking correction, the residual OPDs are attenuated below the spectral density:
\begin{equation}
{\rm PSD_{ \vec  \Phi_{n}}}(f<10\,Hz) \leq 0.001 \, \mu{\rm m}^2/{\rm Hz}.
\end{equation}

The fringe tracker is therefore clearly a high-pass filter. The cutoff frequency of the fringe tracker correction is not well defined because it depends on the efficiency of the Kalman filter and the accuracy with which the equations of state reflect the system. 
On the UTs, the 3\,dB cutoff frequency is on the order of 10\,Hz. On the ATs, maybe because the vibrations are less frequent and the predictive model more accurate, the fringe tracker performance has a higher cutoff frequency at 30\,Hz. This 30\,Hz shows that the system is optimized because it is close to the  cutoff frequency of the open-loop system ($\approx 60\,$Hz in Fig.~\ref{fig:bode}).

\section{Discussion}
\label{sec:conclusion}

\subsection{Have we reached the ultimate sensitivity?}

 It is often assumed that the sensitivity of an optical interferometer must decrease as a function of $N$, the number of telescopes. 
This is not necessarily true. 
The limiting sensitivity is when each of the degrees of freedom of the fringe tracker reaches a variance of one during one coherence time of the atmosphere, 
and each of the degrees of freedom corresponds to a non-zero eigenvalue of  matrix $\mathbb I$ (Section~\ref{sec:I}). Therefore the threshold on the phase-delay control loop ($s/n_{\rm threshold}^{\rm PD}$) is applied on the eigenvalues in Eq.~(\ref{eq:sinvPD}).
For $N>>1$, the signal grows like the flux ($\propto N$), the degrees of freedom like $N$, and the
number of pixels like the number of baselines, $N^2$. For an increasing number of telescopes, the detector noise therefore becomes ever stronger. 
However, in the case of a system without background and detector noise, the S/N on the eigenvalues no longer depends on $N$  \citep{2017A&A...602A.116W}.
The ultimate sensitivity of the fringe tracker can then be established at one photon per coherence time and per telescope, regardless of the number of telescopes. 
For the VLTI, assuming 8-meter telescopes, a coherence time of 10\,ms, a throughput of 1\,\%, and using the full K band, the ultimate sensitivity is Kmag$=17.5$. Even assuming the need for an S/N of 1.5 per baseline and per coherence time, we should be able to reach a Kmag of 16 (Fig.~\ref{fig:SNR}).

What can be done?
There are several paths forward to reach this magnitude. The first is to decrease the background and detector noise. 
The sky brightness in the K band at Paranal is about 12.8 mag/square arcsec\footnote{Table 6 in \url{https://www.eso.org/observing/etc/doc/skycalc/The\_Cerro\_Paranal\_Advanced\_Sky\_Model.pdf}}. 
Therefore, the fraction of sky background light entering the 60\,mas single-mode fiber is almost negligible (${\rm Kmag}_{\rm sky} \leq 19$). A more important light emitter is the VLTI thermal background because of the $\approx 75\%$ absorption of the optical surfaces. All of these surfaces are typically about T = 283\,K to 288\,K and contribute to the majority of the background light. Regarding the detector noise, even if the SAPHIRA detector has a readout noise below 1\,e$^-$, the number of pixels used by the fringe tracker per telescope is 36 (72 in split polarizations). 
For faint objects, this noise can therefore dominate. 
Solutions to use fewer pixels have been proposed \citep{2014SPIE.9146E..2PP,2016SPIE.9907E..1FP} and could lead to a better sensitivity. 
The development of infrared detectors could also be a promising path forward.

The second path is to increase the coherent throughput. On average, only 1\% of the total flux reaches the fringe tracker detector. The  20 mirrors between M1 and the GRAVITY cryostat absorb up to 75\% of the light. In addition, 50\% of the light is scattered inside the IO beam combiner, 50\% is lost by using the beamsplitter on the on-axis mode, and 50\% because of the amplification of the detector (the so-called multiplicative noise).
The last part ($\approx 60\%$ of the remaining flux) is lost when the light is injected into the single-mode fiber.
Throughput improvements could therefore come from  using fewer optical elements and possibly switching more of the mirrors involved from aluminium to gold coatings. Focus can also be placed on a more efficient light coupling into the single-mode fiber. 
This would mean a better AO system for a better Strehl ratio. The advantages of  a good Strehl ratio are manyfold. It increases the mean throughput. It also maximizes the fringe contrast by giving a better instantaneous flux equilibrium. Last, it avoids flux drop-out, which is detrimental for good fringe tracking and model prediction.

Third, but not least, special care will be taken in monitoring and removal of the vibrations. The vibrations cause two main problems. Because they are usually at high frequency, they are difficult to predict and affect the fringe-tracking performance (in contrast to the atmospheric perturbations, which are easier to correct because they are at a lower frequency). The main problem, however, is that they limit how slow the fringe tracker can run because the decrease in fringe contrast hurts more than the additional integration time.

\subsection{Have we reached the ultimate accuracy?}

A false assumption is that sensitivity can be gained by trading accuracy for sensitivity because coherencing (i.e., keeping the fringes within the coherent length) requires fewer photons per coherence time than fringe tracking. However, by simultaneously using the group delay and the phase delay, it is possible to have the best of both worlds: the group-delay controller does the coherencing, while the phase-delay controller works in parallel as far as the S/N permits (Fig.~\ref{fig:OPDrms_snr}).

Despite the high sensitivity, we routinely track high S/N fringes within 100\,nm residual rms with GRAVITY.
However, when the coherence time is short ($\tau_0<3\,$ms at 500\,nm), the fringe tracker performance degrades  (Fig.~\ref{fig:OPDrms}). This is caused by the open-loop latency of the
fringe tracker (of about 4\,ms). Observing during these conditions could clearly benefit from a fringe tracker with a shorter response time. Could we still improve the control loop during good atmospheric conditions, however?

The answer is yes. It lies in the proper management of the S/N by the Kalman filter. 
For convenience, we used an asymptotic estimation of the Kalman gain from the Riccati equation.
A better Kalman filter would propagate errors as well as the state by also applying the equation of state to the covariance matrix:
\begin{equation}
\hat \Sigma_{ \vec x,n|n-1}= \mathsf  A_V \cdot  \hat \Sigma_{ \vec x,n-1} \cdot \mathsf  A_V^\top
,\end{equation}
and derive the optimum gain each DIT. This could be achieved with additional computing power.

An additional amelioration could come from modal control. This was proposed in  \citet{2012A&A...541A..81M} and could theoretically be implemented. 
However, we have currently not been able to find any practical implementation that would make it robust for a realistic environment.

\begin{acknowledgements}
GRAVITY was developed in a
collaboration of the Max Planck Institute for Extraterrestrial Physics,
LESIA of Paris Observatory, IPAG of Université Grenoble Alpes / CNRS,
the Max Planck Institute for Astronomy, the University of Cologne, the
Centro Multidisciplinar de Astrofisica from Lisbon and Porto, and the European
Southern Observatory.
SL would like to thank C.Kulcs\'ar, H.-F.~Raynaud, K.~Houairi, J.~Lozi, J.-M.~Conan, and L.~Mugnier for their insight and remarkable discussions.
A special thanks goes to the pioneering work made by M.~Colavita. SL was supported by the European Union under ERC grant 639248 LITHIUM, and FC by ONERA internal funding.
GP and KP acknowledge support from Action Sp\'ecifique ASHRA of CNRS/INSU and CNES, 
Commission Spécialisée Astronomie-Astrophysique of CNRS/INSU, Conseil Scientifique of Observatoire de Paris and Observatoire des Sciences de l’Univers de Grenoble. AA acknowledges funding from Fundação para a Ciência e Tecnologia through grants PTDC/CTE-AST/116561/2010 and  UID/FIS/00099/2013.
Additional thanks go to Sylvain Rousseau for spoting several typos (eg. Eq 18, 28, 31).
\end{acknowledgements}

%
   \bibliographystyle{aa} 
   \bibliography{FTbib} 
%
%
%

\end{document}